\documentclass[preprint,pre,aps,showpacs,showkeys,amsmath]{revtex4}
\usepackage{graphicx}

\begin{document}

\title{Emergence of Sparsely Synchronized Rhythms and Their Responses to External Stimuli in An Inhomogeneous Small-World Complex Neuronal Network}

\author{Sang-Yoon Kim}
\email{sykim@icn.re.kr}
\author{Woochang Lim}
\email{wclim@icn.re.kr}
\affiliation{Institute for Computational Neuroscience and Department of Science Education, Daegu National University of Education, Daegu 42411, Korea}

\begin{abstract}
By taking into consideration the inhomogeneous population of interneurons in real neural circuits, we consider an inhomogeneous small-world network (SWN) composed of inhibitory short-range (SR) and long-range (LR) interneurons, and investigate the effect of network architecture on emergence of sparsely synchronized rhythms by varying the fraction of LR interneurons $p_{long}$. The betweenness centralities of the LR and the SR interneurons (characterizing the potentiality in controlling communication between other interneurons) are distinctly different, although they have the same average in- and out-degrees (representing the potentiality in communication activity). Hence, in view of the betweenness, SWNs we consider are inhomogeneous, unlike the ``canonical'' Watts-Strogatz SWN with nearly same betweenness centralities. For small $p_{long}$, the average betweenness centrality of LR interneurons is much larger than that of SR interneurons. Hence, the load of communication traffic is much concentrated on a few LR interneurons. However, with further increase in $p_{long}$ the number of LR connections (coming from LR interneurons) increases, and then the average betweenness centrality of LR interneurons decreases. Consequently, the average path length becomes shorter, and the load of communication traffic is less concentrated on LR interneurons, which leads to better efficiency of global communication between interneurons. Sparsely synchronized rhythms are thus found to emerge when passing a small critical value $p_{long}^{(c)}$ $(\simeq 0.16)$. This transition from desynchronization to sparse synchronization is well characterized in terms of a realistic ``thermodynamic'' order parameter, and the degree of sparse synchronization is well measured in terms of a realistic ``statistical-mechanical'' spiking measure. These dynamical behaviors in the inhomogeneous SWN are also compared with those in the homogeneous Watts-Strogatz SWN, in connection with their network topologies. Particularly, we note that the main difference between the two types of SWNs lies in the distribution of betweenness centralities. Unlike the case of the Watts-Strogatz SWN, dynamical responses to external stimuli vary depending on the type of stimulated interneurons in the inhomogeneous SWN. We consider two cases of external time-periodic stimuli applied to sub-populations of LR and SR interneurons, respectively. Dynamical responses (such as synchronization suppression and enhancement) to these two cases of stimuli are studied and discussed in relation to the betweenness centralities of stimulated interneurons, representing the effectiveness for transfer of stimulation effect in the whole network.
\end{abstract}


\maketitle

\section{Introduction}
\label{sec:INT}
A neural circuit in the major parts of the mammalian brain consists of excitatory principal cells and inhibitory interneurons. These interneurons are very diverse in their morphologies and functions, in contrast to the more homogeneous population of principal cells \cite{Buz1}. A majority of short-range (SR) interneurons with mainly local connections coordinate multiple operations in principal cells, while a smaller fraction of long-range (LR) interneurons, with their axons distributed over large areas, innervate and coordinate all interneuron classes for generation of global synchrony in interneuronal networks \cite{Buz,LR1,LR2,LR3,LR4,LR5,LR6}. By providing a coherent oscillatory output to the principal cells, interneuronal networks play the role of the backbones of many brain rhythms \cite{Buz1,Buz,Wang,W_Review}. Here, we are interested in fast sparsely synchronized brain rhythms, associated with diverse cognitive functions (e.g., sensory perception, feature integration, selective attention, and memory formation) \cite{W_Review}. At the population level, synchronous fast oscillations (e.g., gamma rhythm (30-100 Hz) during awake behaving states and rapid eye movement sleep and sharp-wave ripple (100-200 Hz) during quiet sleep and awake immobility) have been observed in local field potential recordings, while at the cellular level individual neuronal recordings have been found to exhibit stochastic and intermittent spike discharges like Geiger counters \cite{SS1,SS2,SS3,SS4,SS5,SS6,SS7}. Thus, for the case of sparsely synchronized rhythms, single-cell firing activity differs distinctly from the population oscillatory behavior, in contrast to fully synchronized rhythms where individual neurons fire regularly at the population frequency like clock oscillators. Under the balance between strong external excitation and strong recurrent inhibition, fast sparse synchronization was found to occur in neuronal networks for both cases of random coupling \cite{Sparse1,Sparse2,Sparse3,Sparse4} and global coupling \cite{Sparse5,Sparse6}. However, in a real neural circuit, synaptic connections are known to have complex topology which is neither regular nor completely random \cite{Buz1,Sporns,CN1,CN2,CN3,CN4,CN5,CN6,CN7}.

Here, we take into consideration the inhomogeneous population of interneurons in real neural circuits, and consider an inhomogeneous small-world network (SWN) composed of two sub-populations of SR and LR interneurons: a preliminary computer modeling was done \cite{Buz1,W_Review,Buz2}. By varying the fraction of LR interneurons $p_{long}$, we study emergence of sparsely synchronized rhythms. As one of complex networks, the Watts-Strogatz SWN, which interpolates between regular lattice (corresponding to the case of $p_{wiring}=0$) with high clustering and random graph (corresponding to the case of $p_{wiring}=1$) with short path length via random uniform rewiring with the probability $p_{wiring}$, has been much studied \cite{SWN1,SWN2,SWN3}. This ``canonical'' Watts-Strogatz SWN is statistically homogeneous because individual interneurons have nearly same topological network centralities such as degree and betweenness. By increasing the rewiring probability $p_{wiring}$, LR connections come uniformly at random from all interneurons over the entire network, in contrast to real neural circuits where such LR connections are not known for the case of SR interneurons \cite{Buz1}. On the other hand, LR connections appear non-uniformly from LR interneurons in the inhomogeneous SWN we consider. For small $p_{long}$, LR interneurons have higher betweenness centralities $B$ (characterizing the potentiality in controlling communication between other interneurons) than SR interneurons \cite{Bet1,Bet2}, although they have the same average in- and out-degrees (representing the potentiality in communication activity). Hence, the load of communication traffic is much concentrated on a few LR interneurons. Then, the LR interneurons  tend to become overloaded by the communication traffic passing through them. For this case, it becomes difficult to get efficient communication between interneurons due to destructive interference between so many signals passing through the LR interneurons \cite{Bet3,SFN-Kim}. However, as $p_{long}$ is further increased, the average betweenness centrality $\left< B \right>_{LR}$ of LR interneurons decreases due to increase in the number of LR interneurons (i.e., because the increased number of LR interneurons share the load of communication traffic). In this way, as a result of increase in $p_{long}$, the average path length $L_p$ (i.e., typical separation between two interneurons represented by average number of synapses between two interneurons along the minimal path) becomes shorter, and the load of communication traffic is less concentrated on LR interneurons because of their decreased average betweenness centrality $\left< B \right>_{LR}$, which leads to better efficiency of global communication between interneurons. Many recent works on various subjects of neurodynamics have been done in SWNs with predominantly local connections and rare long-distance connections \cite{CN6,SW2,SW3,SW4,SW5,SW6,SW7,SW8,SW9,SW10,SW11,SW12,SW13,SWN-Kim}.

In this paper, we study the effect of network topology such as the average path length and the betwenness centrality on emergence of fast sparsely synchronized rhythm by varying $p_{long}$ in the inhomogeneous SWN. For $p_{long} = 0$, the average path length $L_p$  is very long because there exist only SR connections, and hence global communication between distant interneurons becomes ineffective. Consequently, an unsynchronized population state appears. However, as $p_{long}$ is increased from 0, LR connections that connect distant interneurons begin to appear non-uniformly from LR interneurons, and then the average path length $L_p$ can be dramatically decreased only by a few short-cuts. However, in the presence of only a few LR interneurons, the load of communication traffic is much concentrated on the LR interneurons with higher betweenness centralities. With further increase in $p_{long}$, the average betweenness centrality $\left< B \right>_{LR}$ of LR interneurons decreases because the increased number of interneurons share the load of communication traffic. Consequently, the load of communication traffic becomes less concentrated on the LR interneurons. Thus, global effective communication between distant interneurons may be available via LR connections. Eventually, when $p_{long}$ passes a small critical value $p_{long}^{(c)}$ $(\simeq 0.16)$, fast sparsely synchronized population rhythm emerges in the whole population because efficiency of global communication between interneurons becomes enough for occurrence of population synchronization thanks to shorter average path length $L_p$ and smaller average betweenness centrality $\left< B \right>_{LR}$ of LR interneurons. This transition from desynchronization to fast sparse synchronization is well described in terms of a realistic ``thermodynamic'' order parameter, based on the instantaneous population spike rate (IPSR) kernel estimate $R(t)$ \cite{RM}. Population synchronization emerges when $R$ oscillates. Furthermore, through calculation of a spatial correlation function between interneurons \cite{SWN-Kim}, this type of global sparse synchronization is found to occur because dynamical correlation length between neuronal pairs covers the whole system thanks to sufficient number of LR connections. For $p_{long} > p_{long}^{(c)}$, the IPSR kernel estimate $R(t)$ oscillates with fast population frequency of 139 Hz, while individual interneurons discharge stochastic spikes at low rates $(\sim 31$ Hz). This fast sparse synchronization is well characterized in terms of a realistic ``statistical-mechanical'' spiking measure, based on $R$, by taking into consideration the occupation and the pacing degrees of spikes in the raster plot of spikes \cite{RM}. With increasing $p_{long}$ from $p_{long}^{(c)}$, the degree of fast sparse synchronization increases monotonically, but its increasing rate becomes slower due to appearance of sufficient number of LR connections. On the other hand, the total (axon) geometric wiring length continues to increase linearly with respect to $p_{long}$. Hence, at an optimal value $p_{long}^{(o)}$ $(\simeq 0.27)$, an optimal fast sparse synchronization is found to emerge through trade-off between synchronization degree and wiring cost in an economic inhomogeneous SWN. These dynamical behaviors in the inhomogeneous SWN are also compared with those in the homogeneous Watts-Strogatz SWN, in connection with their network topologies. Particularly, we note that the distributions of betweenness centralities are distinctly different in both types of SWNs. In contrast to the case of the Watts-Strogatz SWN, dynamical responses to external stimuli are found to vary depending on the type of stimulated interneurons in the inhomogeneous SWN. We consider two cases of external time-periodic stimuli applied to sub-populations of LR and SR interneurons, respectively. Dynamical responses to these two cases of stimuli are studied and discussed in relation to the betweenness centralities of stimulated interneurons, representing the effectiveness for transfer of stimulation effect in the whole network. It is thus found that the degree of dynamical response (e.g., synchronization suppression or enhancement) for the case of stimulated LR interneurons with higher betweenness centralities is larger (i.e., more suppressed or enhanced) than that for the case of stimulated SR interneurons with lower betweenness centralities.

This paper is organized as follows. In Sec.~\ref{sec:ISWN}, we describe an inhomogeneous SWN composed of SR and LR fast spiking interneurons, and then the governing equations for the population dynamics are given. Then, in Sec.~\ref{sec:Main} we study the effect of network topology on emergence of fast sparsely synchronized rhythms by varying $p_{long}$. Furthermore, we also investigate dynamical responses to external time-periodic stimuli (applied to sub-populations of LR and SR interneurons, respectively) in connection with the betweenness centralities $B$ of stimulated interneurons in Sec.~\ref{sec:DR}. Finally, in Sec.~\ref{sec:SUM}, a summary is given. Explanations on methods for topological and dynamical characterizations are also made in Appendices ~\ref{sec:NTG} and \ref{sec:Dynamics}, respectively.

\section{Inhomogeneous SWN Composed of SR and LR Fast-Spiking Interneurons}
\label{sec:ISWN}
In this section, we first describe an inhomogeneous directed SWN composed of two sub-populations of SR and LR interneurons in the subsection \ref{subsec:ISWN}. Then, the governing equations for the population dynamics in the inhomogeneous SWN are given in the subsection \ref{subsec:GE}.

\subsection{Inhomogeneous SWN}
\label{subsec:ISWN}
\begin{figure}
\includegraphics[width=0.65\columnwidth]{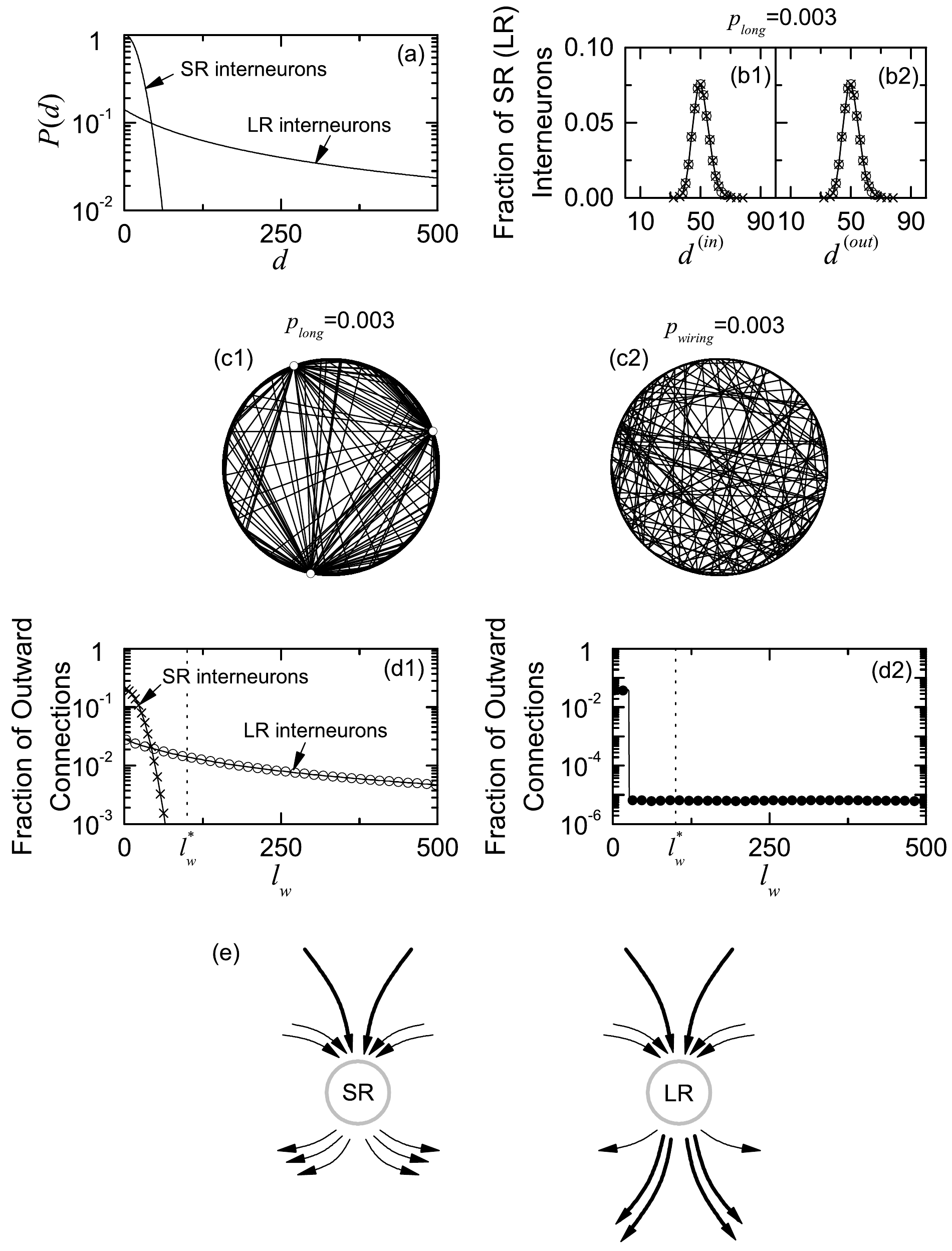}
\linespread{1.1}
\caption{Inhomogeneous SWN composed of SR and LR interneurons. (a) Plots of Gaussian and power-law connection probabilities $P(d)$ versus the distance $d$ between interneurons for the SR and the LR interneurons, respectively. Histograms for (b1) in- and (b2) out-degrees of the SR (crosses) and the LR (open circles) interneurons in the inhomogeneous SWN for $p_{long}=0.003$. Histograms in (b1) and (b2) are obtained through 300 realizations, and the bin size for the histograms is 1. Synaptic connections for $N=10^3$ (total number of interneurons) in (c1) the inhomogeneous SWN for $p_{long}=0.003$ (three LR interneurons are denoted by open circles) and (c2) the homogeneous Watts-Strogatz SWN for $p_{wiring}=0.003$. Histograms for axonal wiring lengths $l_w$ of outward connections of (d1) the SR (crosses) and the LR (open circles) interneurons in the inhomogeneous SWN for $p_{long}=0.003$ and (d2) the interneurons (solid circles) in the homogeneous Watts-Strogatz SWN for $p_{wiring}=0.003$. Histograms for the SR and the LR interneurons in (d1) are obtained through $10^3$ realizations, the bin sizes for the histograms are 1, and $l_w^* (=100)$  is the threshold value for determining whether connections are SR or LR. Fitted Gaussian and power-law solid curves in (d1) agree well with numerically-obtained data for both cases of SR and LR interneurons, respectively. The number of realizations for the histogram in (d2) is $10^3$, and the bin size for the histogram is 1. (e) Schematic representation of the inward and the outward connections of the SR and the LR interneurons.
Long heavy solid lines denote LR connections, while others represent SR connections.
}
\label{fig:ISWN}
\end{figure}

We consider an inhomogeneous directed SWN composed of $N$ inhibitory SR and LR interneurons equidistantly placed on a one-dimensional ring of radius $N/ 2 \pi$. The fraction of LR interneurons may be controlled by the parameter
$p_{long}$ ($ 0 \leq p_{long} \leq 1$). The connection probabilities $P(d)$ from the SR and the LR interneurons are shown in Fig.~\ref{fig:ISWN}(a). For the case of SR interneurons, the connection probability $P(d)$ from each SR interneuron to other interneurons is given by the narrow Gaussian distribution:
\begin{equation}
 P(d) = e^{-d/2 \sigma^2},
\label{eq:Gaussian}
\end{equation}
where $d$ represents the distance between interneurons, and $\sigma$ denotes the standard deviation.
On the other hand, LR interneurons are connected to other interneurons with a (slowly-decaying) power-law distribution:
\begin{equation}
P(d) = A (d+\kappa)^{-\alpha}.
\label{eq:Power}
\end{equation}
Here, we consider the case of $\alpha=1$, and the coefficient $A$ $[=\sqrt{\pi/2} \sigma / (\ln{(N/2+\kappa)} - \ln \kappa)]$ is determined such that the average in- and out-degrees of the SR and the LR interneurons are the same.
Figures \ref{fig:ISWN}(b1) and \ref{fig:ISWN}(b2) show the histograms for the in- and out-degrees of the SR and the LR interneurons for $p_{long}=0.003$, respectively, when $\sigma=20$ and $\kappa=100$. As shown well in the narrow bell-shaped distributions, the SR (crosses) and the LR (circles) interneurons have the same average in- and out-degrees whose values are 50. For comparison, we also consider the Watts-Strogatz SWN with the same average number of
synaptic inputs $M_{syn} (=50)$. LR connections in the inhomogeneous SWN come non-uniformly out of the LR interneurons, in contrast to the Watts-Strogatz SWN where LR connections come uniformly from all the interneurons, as shown
in Figs.~\ref{fig:ISWN}(c1) and \ref{fig:ISWN}(c2) for $p_{long}=p_{wiring}=0.003$. Figure \ref{fig:ISWN}(d1) shows histograms for axonal wiring lengths $l_w$ of outward connections of the SR (crosses) and the LR (circles) interneurons for $p_{long}=0.003$. Here, we set the threshold value $l_w^*$ (=100) for determining whether or not connections are LR, which corresponds to be twice as much as the average in- and out-degrees (=50). As can be expected from the connection distributions $P(d)$, the distributions for wiring lengths $l_w$ of outward connections for the SR (crosses) and the LR (circles) interneurons are well fitted with the narrow Gaussian solid curve [$A e^{-l_w^2/2 \sigma^2};$ $A=0.2$ and $\sigma=20$] and the slowly-decreasing power-law solid curve [$A (l_w +100)^{-1}$; $A=2.8$], respectively. Hence, for the SR interneurons, only SR connections appear, while in the case of LR interneurons, about 61 \% of the total outward connections are LR connections. For comparison, in the case of the homogeneous Watts-Strogatz SWN with $p_{wiring}=0.003$, the histogram for axonal wiring lengths $l_w$ of outward connections of the interneurons is also given in Fig.~\ref{fig:ISWN}(d2). The Watts-Strogatz SWN interpolates between the regular lattice and the random network via random uniform rewiring \cite{SWN1}. For the case of regular network, each interneuron is connected to its first $M_{syn}$ (=50) neighbors ($M_{syn}/2$ on either side) on a ring via outward connections, and then rewire each outward connection uniformly at random over the entire ring with the probability $p_{wiring}$ such that self-connections and duplicate connections are excluded. We note that such random uniform rewiring (independent of the distance between interneurons) is in contrast to the connection probabilities (dependent on the distance) for the SR and the LR interneurons in the inhomogeneous SWN. For $p_{wiring}=0$, the wiring lengths of the total 50000 outward connections are distributed with equal fraction (=0.04) in the range of $1 \leq l_w \leq 25$; otherwise, the fraction is zero. For $p_{wiring}=0.003$, 0.3 \% of the total connections (=150) are rewired uniformly at random in the range of $1 \leq l_w \leq 500$ without self-connections and duplicate connections. However, the possibility that wiring lengths of the rewired connections lie in the range of $1 \leq l_w \leq 25$ is negligibly small, because the value of $p_{wiring}$ is very small. Almost of them are equally distributed in the range of $26 \leq l_w \leq 500$. Hence, the equal fraction for the wiring lengths of the outward connections in the range of $1 \leq l_w \leq 25$ is reduced to 0.0399, while the wiring lengths of the rewired connections are equally distributed with the fraction ($\simeq 6.32 \times 10^{-6}$) in the range of $26 \leq l_w \leq 500$.

For illustrative purpose, an example of the inward and the outward connections for the SR and the LR interneurons is shown in Fig.~\ref{fig:ISWN}(e). In this example, the average in- and out-degrees (=6) are the same for both the SR and the LR interneurons. For the case of outward connections, the SR interneuron has no LR connections, while some connections (=4) of the LR interneuron are LR connections (denoted by long heavy solid lines). However, in the case of inward connections, there is no particular difference between the SR and the LR interneurons (i.e., LR and SR inward connections may appear for both the SR and the LR interneurons). Aa a result, the mean wiring length $\overline{l_w}$ of outward connections for the LR interneuron is longer than that for the SR interneuron, which might lead to some difference in the betweenness centralities $B$ of the SR and the LR interneurons [e.g., refer to Fig.~\ref{fig:BT}(b)].

\subsection{Governing Equations for The Population Dynamics}
\label{subsec:GE}
As an element in our SWN, we choose the fast spiking (FS) Izhikevich interneuron model which is not only biologically plausible, but also computationally efficient \cite{Izhi1,Izhi2,Izhi3,Izhi4}.
Then, we consider an SWN composed of $N$ SR and LR interneurons, where $N$ is the total number of interneurons. The fraction of LR interneurons among the $N$ interneurons
is given by the parameter $p_{long}$. The connection probability $P(d)$ of the interneurons depend on their type (i.e., it depends on whether the interneurons are LR or SR), as given in Eqs.~(\ref{eq:Gaussian}) and
(\ref{eq:Power}). The following equations (\ref{eq:PD1})-(\ref{eq:PD7}) govern the population dynamics in the SWN:
\begin{eqnarray}
C\frac{dv_i}{dt} &=& k (v_i - v_r) (v_i - v_t) - u_i +I_{DC} +D \xi_{i} -I_{syn,i} + S_i(t), \label{eq:PD1} \\
\frac{du_i}{dt} &=& a \{ U(v_i) - u_i \},  \;\;\; i=1, \cdots, N, \label{eq:PD2}
\end{eqnarray}
with the auxiliary after-spike resetting:
\begin{equation}
{\rm if~} v_i \geq v_p,~ {\rm then~} v_i \leftarrow c~ {\rm and~} u_i \leftarrow u_i + d, \label{eq:PD3}
\end{equation}
where
\begin{eqnarray}
U(v) &=& \left\{ \begin{array}{l} 0 {\rm ~for~} v<v_b \\ b(v - v_b)^3 {\rm ~for~} v \ge v_b \end{array} \right. , \label{eq:PD4} \\
I_{syn,i} &=& \frac{J}{d_i^{(in)}} \sum_{j=1 (\ne i)}^N w_{ij} s_j(t) (v_i - V_{syn}), \label{eq:PD5}\\
s_j(t) &=& \sum_{f=1}^{F_j} E(t-t_f^{(j)}-\tau_l);~E(t) = \frac{1}{\tau_d - \tau_r} (e^{-t/\tau_d} - e^{-t/\tau_r}) \Theta(t), \label{eq:PD6} \\
S_i(t) &=& \alpha_i A \sin(\omega_d t). \label{eq:PD7}
\end{eqnarray}
Here, $v_i(t)$ and $u_i(t)$ are the state variables of the $i$th interneuron at a time $t$ which represent the membrane potential and the recovery current, respectively. These membrane potential and the recovery variable, $v_i(t)$ and $u_i(t)$, are reset according to Eq.~(\ref{eq:PD3}) when $v_i(t)$ reaches its cutoff value $v_p$. $C$, $v_r$, and $v_t$ in Eq.~(\ref{eq:PD1}) are the membrane capacitance, the resting membrane potential, and the instantaneous threshold potential, respectively. The parameter values used in our computations are listed in Table \ref{tab:Parm}. More details on the FS Izhikevich interneuron model, the external common stimulus to all the FS interneuron, the synaptic currents, the external time-periodic stimulus to sub-populations of LR and SR interneurons, and numerical integration of the governing equations are given in the following subsubsections.

\subsubsection{FS Izhikevich Interneuron Model}
\label{subsubsec:Izhi}
The Izhikevich model matches neuronal dynamics by tuning the parameters $(k, a, b, c, d)$ instead of matching neuronal electrophysiology, unlike the Hodgkin-Huxley-type conductance-based models \cite{Izhi1,Izhi2,Izhi3,Izhi4}.
The parameters $k$ and $b$ are related to the neuron's rheobase and input resistance, and $a$, $c$, and $d$ are the recovery time constant, the after-spike reset value of $v$, and the total amount of outward minus inward currents during the spike and affecting the after-spike behavior (i.e., after-spike jump value of $u$), respectively. Depending on the values of these parameters, the Izhikevich neuron model may exhibit 20 of the most prominent neuro-computational features of cortical neurons \cite{Izhi1,Izhi2,Izhi3,Izhi4}. Here, we use the parameter values for the FS interneurons in the layer 5 rat visual cortex, which are listed in the 1st item of Table \ref{tab:Parm}.

\subsubsection{External Common Stimulus to The FS Izhikevich Interneuron}
\label{subsubsec:Sti}
Each Izhikevich interneuron is stimulated by both a common DC current $I_{DC}$ and an independent Gaussian white noise $\xi_i$ [see the 3rd and the 4th terms in Eq.~(\ref{eq:PD1})]. The Gaussian white noise satisfies $\langle \xi_i(t) \rangle =0$ and $\langle \xi_i(t)~\xi_j(t') \rangle = \delta_{ij}~\delta(t-t')$, where $\langle\cdots\rangle$ denotes the ensemble average. Here, the Gaussian noise $\xi$ may be
regarded as a parametric one which randomly perturbs the strength of the applied current $I_{DC}$, and its intensity is controlled by the parameter $D$. For $D=0$, the Izhikevich interneuron exhibits a jump from a resting state to a spiking state via subcritical Hopf bifurcation at a higher threshold $I_{DC,h} (\simeq 73.7)$ by absorbing an unstable limit cycle born through a fold limit cycle bifurcation for a lower threshold $I_{DC,l} (\simeq 72.8)$. Hence, the Izhikevich interneuron exhibits type-II excitability because it begins to fire with a non-zero frequency \cite{Ex1,Ex2}. As $I_{DC}$ is increased from $I_{DC,h}$, the mean firing rate $f$ increases monotonically.
The values of $I_{DC}$ and $D$ used in this paper are given in the 2nd item of Table \ref{tab:Parm}.

\subsubsection{Synaptic Currents}
\label{subsubsec:Syn}
The 5th term in Eq.~(\ref{eq:PD1}) denotes the synaptic couplings of inhibitory FS interneurons. $I_{i,syn}$ of Eqs.~(\ref{eq:PD5}) represents the synaptic current injected into the $i$th interneuron.
The synaptic connectivity is given by the connection weight matrix $W$ (=$\{ w_{ij} \}$) where $w_{ij}=1$ if the neuron $j$ is presynaptic to the interneuron $i$; otherwise, $w_{ij}=0$.
Here, the synaptic connection is modeled in terms of the inhomogeneous SWN. Among the $N$ interneurons, LR interneurons are randomly chosen with the fraction of LR interneurons $p_{long}$.
Then, connections form the SR and the LR interneurons to other interneurons are made with the connection probabilities $P(d)$ of Eqs.~(\ref{eq:Gaussian}) and (\ref{eq:Power}), respectively.
Through this process, the connection weight matrix $W$ may be completed. Then, the in-degree of the $i$th neuron, $d_{i}^{(in)}$ (i.e., the number of synaptic inputs to the neuron $i$) is given by $d_{i}^{(in)} =
\sum_{j=1(\ne i)}^N w_{ij}$. For this case, the average number of synaptic inputs per interneuron is $M_{syn} = \frac{1}{N} \sum_{i=1}^{N} d_{i}^{(in)}$.

The fraction of open synaptic ion channels at time $t$ is denoted by $s(t)$. The time course of $s_j(t)$ of the $j$th interneuron is given by a sum of delayed double-exponential functions $E(t-t_f^{(j)}-\tau_l)$ [see Eq.~(\ref{eq:PD6})], where $\tau_l$ is the synaptic delay, and $t_f^{(j)}$ and $F_j$ are the $f$th spiking time and the total number of spikes of the $j$th interneuron at time $t$, respectively. Here, $E(t)$ [which corresponds to contribution of a presynaptic spike occurring at time $0$ to $s(t)$ in the absence of synaptic delay] is controlled by the two synaptic time constants: synaptic rise time $\tau_r$ and decay time $\tau_d$, and $\Theta(t)$ is the Heaviside step function: $\Theta(t)=1$ for $t \geq 0$ and 0 for $t <0$. The synaptic coupling strength is controlled by the parameter $J$,  and $V_{syn}$ is the synaptic reversal potential. For the inhibitory GABAergic synapse (involving the $\rm{GABA_A}$ receptors), the values of $\tau_l$, $\tau_r$, $\tau_d$, and $V_{syn}$ are listed in the 3rd item of Table \ref{tab:Parm}.

For comparison, we also consider the connectivity of the Watts-Strogatz SWN which interpolates between regular lattice (corresponding to the case of $p_{wiring}=0$) and random graph (corresponding to the case of $p_{wiring}=1$)  via random uniform rewiring with the probability $p_{wiring}$ \cite{SWN1,SWN2,SWN3}. For $p_{wiring}=0,$ we start with a directed regular ring lattice with $N$ FS Izhikevich interneurons where each Izhikevich interneuron is coupled to its first $M_{syn}$ neighbors ($M_{syn}/2$ on either side) via outward synapses, and rewire each outward connection uniformly at random over the whole ring with the probability $p_{wiring}$ (without self-connections and duplicate connections). This random uniform rewiring is made independently of the distance between interneurons, in contrast to the connection probabilities (varying depending on the distance) for the SR and the LR interneurons in the inhomogeneous SWN. Here, we set $M_{syn}=50$ which is the same as the average in-degree in the inhomogeneous SWN.

\subsubsection{External Time-Periodic Stimulus}
The last term in Eq.~(\ref{eq:PD1}) represents the external time-periodic stimulus to the $i$th interneuron, $S_i(t)$, the explicit form of which is given in Eq.~(\ref{eq:PD6}). If stimulated to the $i$th interneuron, $\alpha_i=1$; otherwise, $\alpha_i=0.$ (In the absence of external stimulus, $\alpha_i=0$ for all $i$.) The driving angular frequency of the stimulus is $\omega_d$, and its amplitude is $A.$ We apply $S_i(t)$ to two sub-populations of LR and SR interneurons with different betweenness centralities, respectively, and investigate their dynamical responses in connection with the betweenness centralities of stimulated interneurons.

\subsubsection{Numerical Method for Integration}
\label{subsubsec:NM}
Numerical integration of stochastic differential Eqs.~(\ref{eq:PD1})-(\ref{eq:PD6}) is done by employing the Heun method \cite{SDE} with the time step $\Delta t=0.01$ ms. For each realization of the stochastic process, we choose a random initial point $[v_i(0),u_i(0)]$ for the $i$th $(i=1,\dots, N)$ interneuron with uniform probability in the range of $v_i(0) \in (-50,-45)$ and $u_i(0) \in (10,15)$.

\section{Effect of Network Architecture on Emergence of Fast Sparsely Synchronized Rhythms in An Inhomogeneous SWN}
\label{sec:Main}
In this section, we study the effect of network architecture on emergence of fast sparsely synchronized rhythms by varying $p_{long}$ in an inhomogeneous SWN. In Subsec.~\ref{subsec:TC}, we first study the network topology, and then the effect of network topology on emergence of sparsely synchronized rhythms is investigated in Subsec.~\ref{subsec:ENT}.

\subsection{Topological Characterization of An Inhomogeneous SWN}
\label{subsec:TC}
\begin{figure}
\includegraphics[width=0.8\columnwidth]{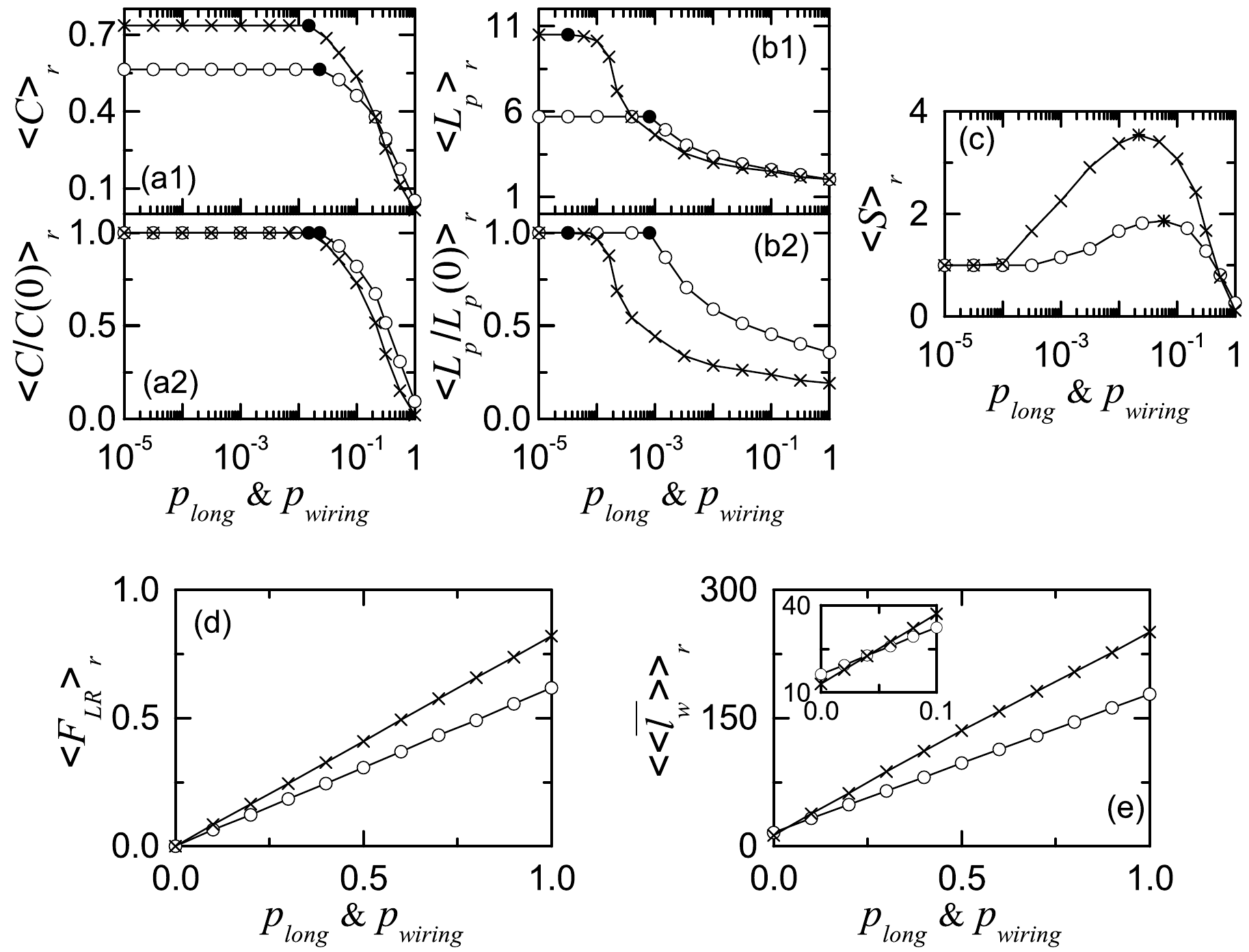}
\caption{Clustering coefficient $C$, average path length $L_p$, and small-worldness coefficient $S$. Circles and crosses denote values for the cases of the inhomogeneous and the homogeneous Watts-Strogatz SWNs with the parameters $p_{long}$ and $p_{wiring}$, respectively. Plots of (a1) the clustering coefficient ${\left< C \right>}_r$ and (a2) the normalized clustering coefficient ${\left< C/C(0) \right>}_r$ versus $p_{long}$ and $p_{wiring}$. Plots of (b1) the average path length ${\left< L_p \right>}_r$  and (b2) the normalized average path length ${\left< L_p/L_p(0) \right>}_r$ versus $p_{long}$ and $p_{wiring}$. The thresholds at which the clustering coefficient $C$ and the average path length $L_p$ begin to decrease rapidly are denoted by solid circles. (c) Plots of the small-worldness coefficient $\left< S \right>_r$ versus $p_{long}$ and $p_{wiring}$; optimal values $p_{long}^*$  and $p_{wiring}^*$  for which $\left< S \right>_r$ has maxima are represented by the stars. $C(0)$  and $L_p(0)$  are values for $p_{long}=p_{wiring}=0$. (d) Plot of the fraction of LR outward connections $F_{LR}$ in the whole population of interneurons versus $p_{long}$ and $p_{wiring}$. (e) Plot of the ensemble-averaged mean wiring length ${\left< \left< \overline{l_w} \right> \right>}_r$  of the outward connections in the whole population of interneurons versus $p_{long}$ and $p_{wiring}$; for more clear presentation, a magnified inset is given for small $p_{long}$ and $p_{wiring}$. $\left< \cdots \right>$ represents the average over all interneurons, and $\left< \cdots \right>_r$ denotes an average over 20 realizations.
}
\label{fig:NT}
\end{figure}

We study the network topology of the inhomogeneous SWN in comparison with that of the Watts-Strogatz SWN in comparison with that of the Watts-Strogatz SWN. Figures \ref{fig:NT}(a1) and \ref{fig:NT}(b1) show plots of the clustering coefficient $C$ and the average path length $L_p$ versus $p_{long}$ and $p_{wiring}$, respectively, in the inhomogeneous SWN (circles) and the Watts-Strogatz SWN (crosses). These values of $C$ and $L_p$ are also normalized by dividing them by the values $C(0)$ and $L_p(0)$ for $p_{long}=p_{wiring}=0$, as shown in Figs.~\ref{fig:NT}(a2) and \ref{fig:NT}(b2), respectively; $C(0) \simeq 0.565~(0.735)$ and $L_P(0) \simeq 5.71~(10.5)$ for the inhomogeneous (Watts-Strogatz) SWN. For $p_{long}= p_{wiring} =0$, both $C$ and $L_p$ in the inhomogeneous SWN are smaller than those in the Watts-Strogatz SWN, and this tendency continues to be valid for sufficiently small $p_{long}$ and $p_{wiring}$. No LR connections appear for both cases of $p_{long}= p_{wiring} =0$ [see Fig.~\ref{fig:NT}(d)].  However, the ensemble-averaged mean wiring length $\left< \overline{l_w)} \right>$ of outward connections coming from the SR interneurons via the Gaussian connection probability for $p_{long}=0$ is longer than that of local connections of interneurons in the regular lattice for $p_{wiring}=0$ [see Fig.~\ref{fig:NT}(e)]; for the case of regular lattice, each interneuron is coupled to its first $M_{syn} (=50)$ neighbors ($M_{syn}/2$ on either side) via outward connections on a ring. As a result of appearance of a little longer connections, both $C$ and $L_p$ become smaller for the case of $p_{long}=0$.

As $p_{long}$ and $p_{wiring}$ are increased from 0, LR connections begin to appear. For the case of inhomogeneous SWN, LR connections come from LR interneurons via the power-law connection probability, while in the case of the Watts-Strogatz SWN, LR connections appear via random uniform rewiring (which is made independently of the distance between interneurons). Consequently, the fraction of LR outward connections (i.e., the ratio of the LR outward connections to the total outward connections) $F_{LR}$ in the Watts-Strogatz SWN is larger than that in the inhomogeneous SWN, as shown in Fig.~\ref{fig:NT}(d). Due to appearance of more LR connections, when passing $p_{long}=p_{wiring}\simeq 0.04$, the ensemble-averaged mean wiring length $\left< \overline{l_w} \right>$ of outward connections for the case of Watts-Strogatz SWN becomes longer than that in the inhomogeneous SWN (i.e., a crossing occurs), as shown in the inset in Fig.~\ref{fig:NT}(e).

With increasing $p_{long}$ and $p_{wiring}$ from 0, LR short-cuts that connect distant interneurons appear. Then, the average path length $L_p$ may be dramatically decreased only by a few short-cuts when passing a lower threshold [$p_{long,l}^{(th)} \simeq 0.0008$ and $p_{wiring,l}^{(th)} \simeq 3 \times 10^{-5}$ which are denoted by solid circles in Figs.~\ref{fig:NT}(b1)], where the clustering coefficient $C$ remains to be high. However, as a higher threshold [$p_{long,h}^{(th)} \simeq 0.023$ and $p_{wiring,h}^{(th)} \simeq 0.015$ which are also represented by solid circles in Fig.~\ref{fig:NT}(a1)] is passed, the cliquishness of a typical neighborhood in the network also begins to break up due to appearance of more LR connections, and then $C$ decreases rapidly. When the two parameters are the same (i.e., $p_{long}=p_{wiring}$), the fraction of LR outward connections $F_{LR}$ for the case of the Watts-Strogatz SWN is larger than that for the case of inhomogeneous SWN [see Fig.~\ref{fig:NT}(d)]. Hence, the effect of LR connections occurs first in the case of the Watts-Strogatz SWN. As a result, the threshold values for the Watts-Strogatz SWN become less than those for the inhomogeneous SWN. Furthermore, when passing crossing values ($p_{long}=p_{wiring}\simeq 0.0004$ for $L_p$ and $p_{long}=p_{wiring} \simeq 0.21$ for $C$), the values of $L_p$ and $C$ for the Watts-Strogatz SWN become smaller than those for the inhomogeneous SWN [i.e., crossings occur between the two curves in Figs.~\ref{fig:NT}(a1) and \ref{fig:NT}(b1)]. After such crossings, due to the effect of continually increased LR connections, the clustering coefficient $C$ continues to decrease rapidly, while the average path length $L_p$ begins to decrease slowly after appearance of sufficient number of LR connections (affecting $L_p$)

An ``optimal'' SWN may appear via trade-off between the high clustering (representing the local efficiency of information transfer) and the short average path length (denoting the reciprocal of the global efficiency of information transfer). To this end, a small-worldness coefficient $S$ is introduced \cite{SWC}:
\begin{equation}
S(p) = \frac{C(p)/C(0)}{L(p)/L(0)}; \;\;\;\; p=p_{long} \;{\rm or}\; p_{wiring}.
\label{eq:SWC}
\end{equation}
Figure \ref{fig:NT}(c) shows plots of the small-worldness coefficient $S$ versus $p_{long}$ and $p_{wiring}$. The maxima of $S$ (denoted by stars) occur at $p_{long}^{(max)} \simeq 0.059$ and $p_{wiring}^{(max)} \simeq 0.022$ for the inhomogeneous SWN and the Watts-Strogatz SWN, respectively. As explained above, the Watts-Strogatz SWN has more LR connections than the inhomogeneous SWN, and hence the effect of LR connections becomes larger for the case of the Watts-Strogatz SWN. Due to the larger effect of LR connections, the maximum for the Watts-Strogatz SWN appears at a smaller parameter value (i.e., $p_{wiring}^{(max)} < p_{long}^{(max)}$), and its maximum value is also larger than that for the inhomogeneous SWN.

\begin{figure}
\includegraphics[width=0.8\columnwidth]{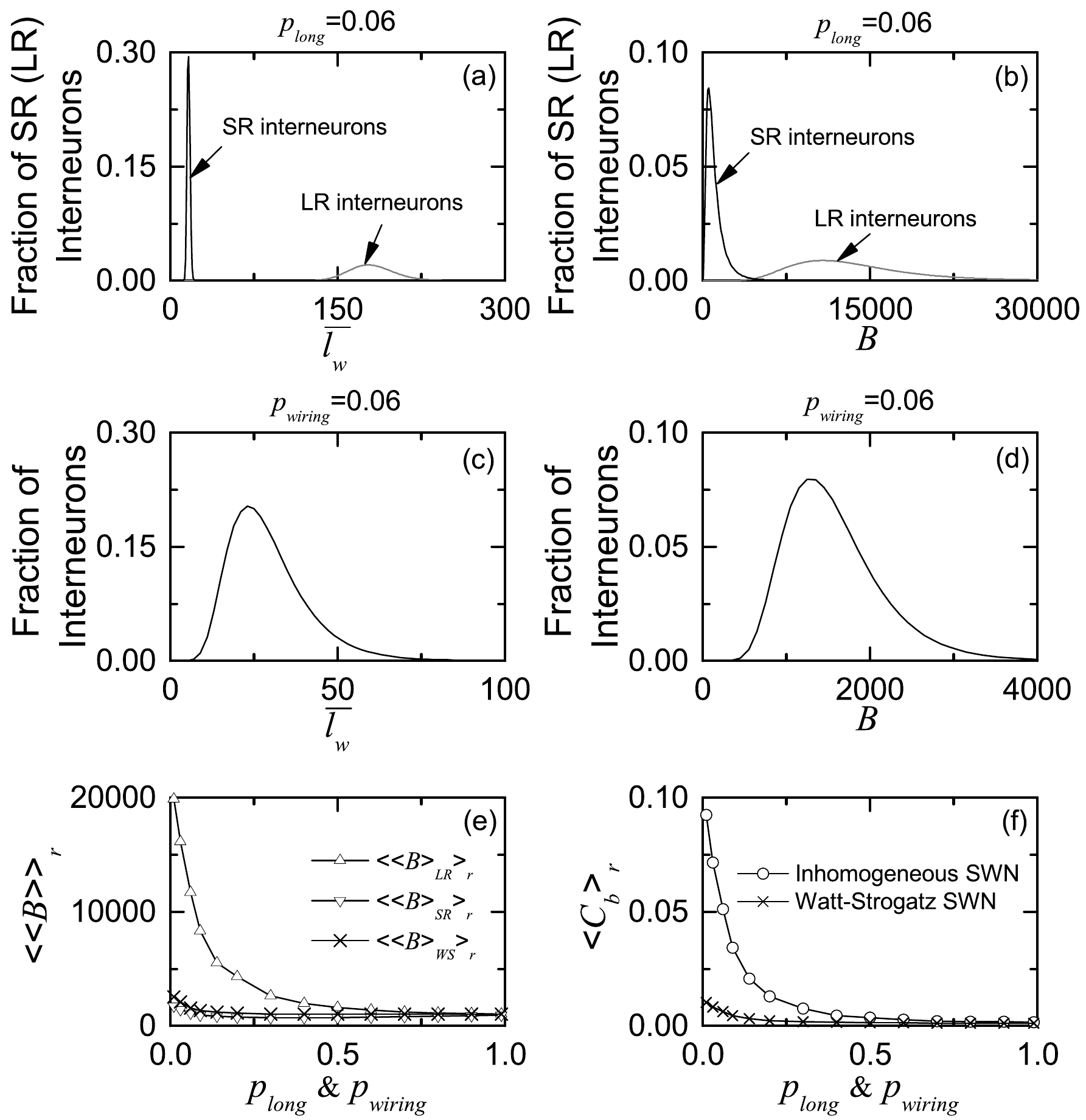}
\caption{Mean wiring length $\overline{l_w}$, betweennes centrality $B$, and betweenness centralization $C_b$. Histograms for (a) the mean wiring length $\overline{l_w}$ of outward connections and (b) the betweenness centrality $B$ of the SR and the LR interneurons in the inhomogeneous SWN for $p_{long}=0.06$. Histograms for (c) the mean wiring length $\overline{l_w}$ of outward connections and (d) the betweenness centrality $B$ of the interneurons in the homogeneous Watts-Strogatz SWN for $p_{wiring}=0.06$. Histograms in (a)-(d) are obtained via 300 realizations. The bin sizes for the histograms in (a)-(d) are 1, 100, 2, and 100, respectively. (e) Plots of average betweenness centrality ${\left< \left< B \right> \right>}_r$ ($\left< \cdots \right>$ represents an ensemble-average over interneurons) versus $p_{long}$ and $p_{wiring}$; upper  and lower triangles denote the average betweenness centralities ($\left< \left< B \right>_{LR} \right>_r$ and $\left< \left< B \right>_{SR} \right>_r$) of the LR and the SR interneurons, respectively, in the inhomogeneous SWN, and  crosses represents the average betweenness centrality $\left< \left< B \right>_{WS} \right>_r$ of the interneurons in the homogeneous Watts-Strogatz SWN. (f) Plot of betweenness centralization $C_b$ versus $p_{long}$ and $p_{wiring}$; circles and crosses represent the values in the inhomogeneous and the homogeneous Watts-Strogatz SWNs, respectively. $\left< \cdots \right>_r$ denotes an average over 20 realizations.
}
\label{fig:BT}
\end{figure}

In the network science, centrality refers to indicators which identify the most important nodes within the network. In Figs.~\ref{fig:ISWN}(b1) and \ref{fig:ISWN}(b2), we consider the simplest degree centrality, which is given by the number of edges of a node. This degree centrality represents the potentiality in communication activity. As explained above, all the SR and LR interneurons in the inhomegeneous SWN have nearly the same in- and out-degrees due to narrow bell-shaped distributions, like the case of the Watts-Strogatz SWN. Betweenness is also another centrality measure of a node within the network. Betweenness centrality $B_i$ of Eq.~(\ref{eq:BT}) for the node $i$ denotes the fraction of all the shortest paths between any two other nodes that pass through the node $i$. Unlike the case of the degree centrality, the distribution of betweenness centralities for the case of the inhomogeneous SWN is inhomogeneous, in contrast to the case of the Watts-Strogatz SWN with nearly same centralities. Figure \ref{fig:BT}(a) shows histograms of the mean wiring length $\overline {l_w}$ of outward connections for both cases of the SR and the LR interneurons for $p_{long}=0.06$. As shown in the figure with two peaks, the ensemble-averaged mean wiring length $\left< \overline {l_w} \right>_{LR}$ ($\simeq 180.6$) over all LR interneurons is much longer than the ensemble-averaged mean wiring length $\left< \overline {l_w} \right>_{SR}$ ($\simeq 16.3$) over all SR interneurons, because LR connections appear non-uniformly from the LR interneurons with the power-law connection probability. Although the mean wiring length is not the only factor affecting the betweenness centrality, it is expected that, in an average sense LR interneurons with longer mean wiring lengths might have larger betweenness centralities. Figure \ref{fig:BT}(b) shows histograms of the betweenness centrality $B$ of the SR and the LR interneuorns for $p_{long}=0.06$. It is thus found that the ensemble-averaged betweenness centrality $\left< B \right>_{LR}$ ($\simeq 13482$) for the case of LR interneurons is much larger than the ensemble-averaged betweenness centrality $\left< B \right>_{SR}$ ($\simeq 1095$) for the case of SR interneurons. In this way, the two sub-populations of the SR and the LR interneurons have distinctly different distributions of the betweenness centralities. For comparison, we also consider the case of the Watts-Strogatz SWN. Figures \ref{fig:BT}(c) and \ref{fig:BT}(d) show the histograms of the mean wiring length $\overline {l_w}$ and the betweenness centrality $B$ for $p_{wiring}=0.06$. Each histogram has only one peak, in contrast to the case of the inhomogeneous SWN with double peaks. Their ensemble-averaged values are $\left< \overline {l_w} \right>_{WS}$ ($\simeq 28$) and $\left< B \right>_{WS}$ ($\simeq 1574$). These ensemble-averaged values are much smaller than those for the case of the LR interneurons, while they are a little larger than those for the case of the SR interneurons. The dispersions of the distributions in the Watts-Strogatz SWN are also smaller than those in the whole population of the inhomogeneous SWN.

By varying the parameters $p_{long}$ and $p_{wiring}$, we also obtain the average betweenness centralities in Fig.~\ref{fig:BT}(e). For sufficiently small $p_{long}$, the average betweenness centrality
$\left< B \right>_{LR}$ (denoted by the upper triangles) of the LR interneurons is much larger than $\left< B \right>_{SR}$ (denoted by the lower triangles) of the SR interneurons. Consequently, load of communication traffic between interneurons is much concentrated on a few LR interneurons with higher betweenness centralities. However, as $p_{long}$ is increased, at first $\left< B \right>_{LR}$ decreases very rapidly, and then its decreasing rate becomes very slow after $p_{long} \sim 0.3$. For the case of the SR interneurons, $\left< B \right>_{SR}$ also decreases a little, but soon it becomes nearly saturated after $p_{long} \sim 0.1$.
For comparison, the average betweenness centralities $\left< B \right>_{WS}$ in the Watts-Strogatz SWN are also represented by crosses, and they are so close to $\left< B \right>_{SR}$ (but, a little larger than $\left< B \right>_{SR}$). For sufficiently large $p_{long}$ and $p_{wiring}$ (larger than about 0.9), LR interneurons become dominant, and they share the load of communication traffic. As a result, the three types of average betweenness centralities become nearly the same (i.e., the load of communication traffic is nearly evenly distributed between interneurons). To examine how evenly the betweenness centrality is distributed among interneurons (i.e., how evenly the load of communication traffic is distributed among interneurons), we consider the group betweenness centralization, representing the degree to which the maximum betweenness centrality $B_{max}$ of the ``head'' LR interneuron exceeds the betweenness centralities of all the other interneurons. The betweenness centralization $C_b$ of Eq.~(\ref{eq:CB}) is given by the sum of differences between the maximum betweenness centrality $B_{max}$ of the head LR interneuron and the betweenness centrality $B_i$ of other interneuron $i$ and normalized by dividing the sum of differences with its maximum possible value. Figure \ref{fig:BT}(f) shows plots of $C_b$ versus $p_{long}$ and $p_{wiring}$ in the inhomogeneous SWN (denoted by circles) and the Watts-Strogatz SWN (represented by crosses). For small $p_{long}$, $C_b$ is large, which implies that load of communication is concentrated on the head LR interneuron, and hence the head LR interneuron tends to become overloaded by the communication traffic passing through it. In this case, it becomes difficult to get efficient communication between interneurons due to destructive interference between so many signals passing through the head LR interneuron \cite{Bet3,SFN-Kim}. However, as $p_{long}$ is increased, $C_b$ decreases very rapidly, and then the load of communication traffic becomes more and more evenly distributed between interneurons. In contrast, for the case of the Watts-Strogatz SWN, $C_b$ is very small even when $p_{wiring}$ is small. With increasing $p_{wiring}$, $C_b$ decreases a little, and soon it becomes nearly saturated. Consequently, load of communication is evenly distributed between interneuonrs in the nearly whole range of $p_{wiring}$.

To summarize the network topology of the inhomogeneous SWN, as $p_{long}$ is increased, the average path length $L_p$ becomes shorter [see Fig.~\ref{fig:NT}(b)] and load of global communication traffic becomes less concentrated
on the head LR interneuron due to decrease in the betweenness centralization $C_b$ (i.e., the load of communication traffic is more evenly distributed between interneurons). Consequently, with increasing $p_{long}$, efficiency of communication between interneurons becomes better, which may result in emergence of sparsely synchronized rhythms, as shown in the next subsection \ref{subsec:ENT}. The inhomogeneous SWN and the Watts-Strogatz SWN have distinctly different distributions of betweenness centralities. Unlike the case of Watts-Strogatz SWN with a single-peaked distribution, the distribution of betweenness centralities in the inhomogeneous SWN is double-peaked: LR interneurons have higher betweenness centralities, while SR interneurons have lower betweenness centralities. Since the betweenness centrality represents the effectiveness of information transfer, the degree of dynamical responses is expected to vary depending on whether external stimuli are applied to LR interneurons or SR interneurons, which will be investigated in Sec.~\ref{sec:DR}.

\subsection{Effect of Network Topology on Emergence of Fast Sparsely Synchronized Rhythms}
\label{subsec:ENT}
\begin{figure}
\includegraphics[width=0.8\columnwidth]{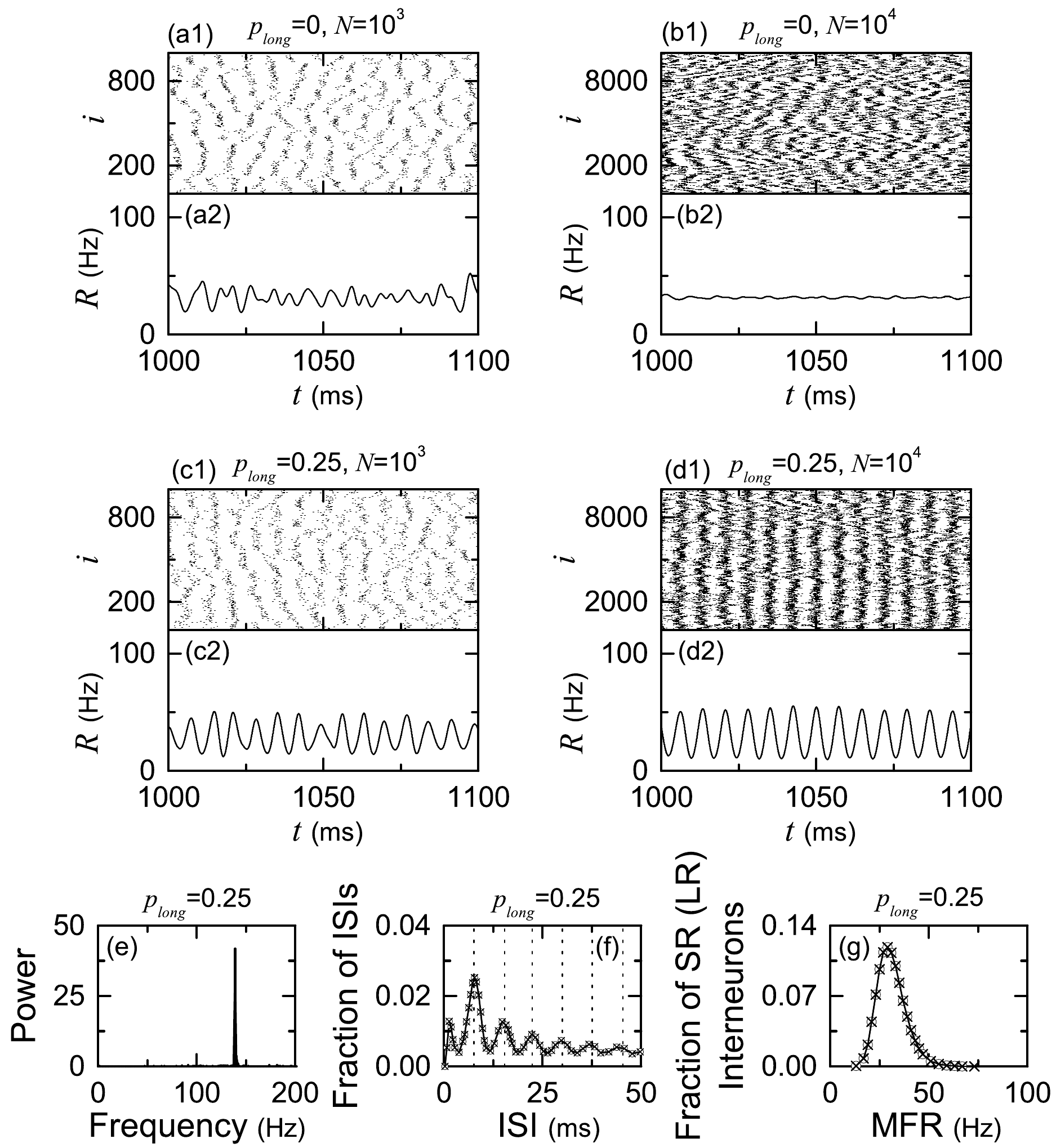}
\linespread{1.1}
\caption{Population states in the inhomogeneious SWN. Unsynchronized state for $p_{long}=0$: raster plots of spikes for $N=$ (a1) $10^3$ and (b1) $10^4$ and IPSR kernel estimates $R(t)$   for $N=$ (a2) $10^3$ and (b2) $10^4$. Synchronized state for $p_{long}=0.25$ : raster plots of spikes for $N=$ (c1) $10^3$ and (d1) $10^4$ and IPSR kernel estimates $R(t)$ for $N=$ (c2) $10^3$ and (d2) $10^4$. The band width of the Gaussian kernel estimate for the IPSR $R(t)$  is 1 ms. In (e)-(g), $p_{long}=0.25$  and  $N=10^3$. (e) One-sided power spectrum of $\Delta R(t) [=R(t) - \overline{R(t)}]$ (the overbar represents the time average) with mean-squared amplitude normalization. The power spectrum is obtained from $2^{16}$ (=65536) data points. (f) Interspike interval (ISI) histograms of the SR and the LR interneurons. Vertical dotted lines denote integer multiples of global period $T_G (\simeq 7.2$ ms) of $R(t)$. Open circles and crosses represent ISIs for the LR and the SR interneurons, respectively. For each interneuron, $5 \times 10^4$ ISIs are obtained through 50 realizations. Hence, the ISI histograms are composed of $5 \times 10^7$ ISIs, and the bin size for the histograms is 0.5 ms. (g) Histograms for the mean firing rates (MFRs) of individual SR and LR interneurons. Open circles and crosses represent MFRs for the LR and the SR interneurons, respectively. Averaging time for the MFR of each interneuron is $10^4$ ms in each realization, it is obtained via 50 realizations, and the bin size for the histogram is 0.5 Hz.
}
\label{fig:PS}
\end{figure}

\begin{figure}
\includegraphics[width=0.65\columnwidth]{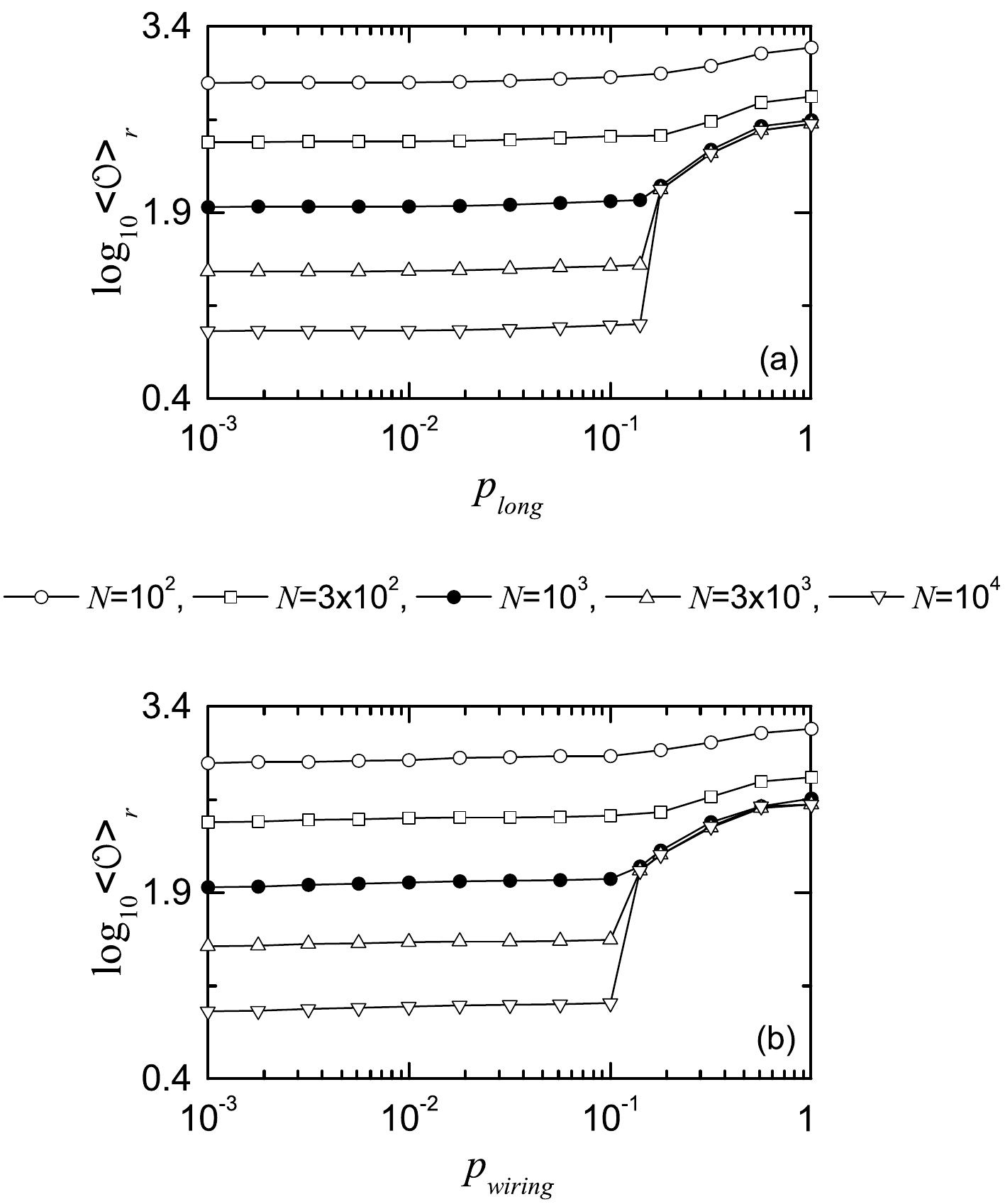}
\caption{Transitions from desynchronization to synchronization in the inhomogeneous and the homogeneous Watts-Strogatz SWNs. Plots of the thermodynamic order parameter $ \log_{10} \left< \cal{O} \right>_r$ versus (a) $p_{long}$ and (b) $p_{wiring}$. Averaging time for $\cal{O}$ is $3 \times 10^4$ ms in each realization, and $\left< \cdots \right>_r$ denotes an average over 20 realizations.
}
\label{fig:Order}
\end{figure}

\begin{figure}
\includegraphics[width=0.7\columnwidth]{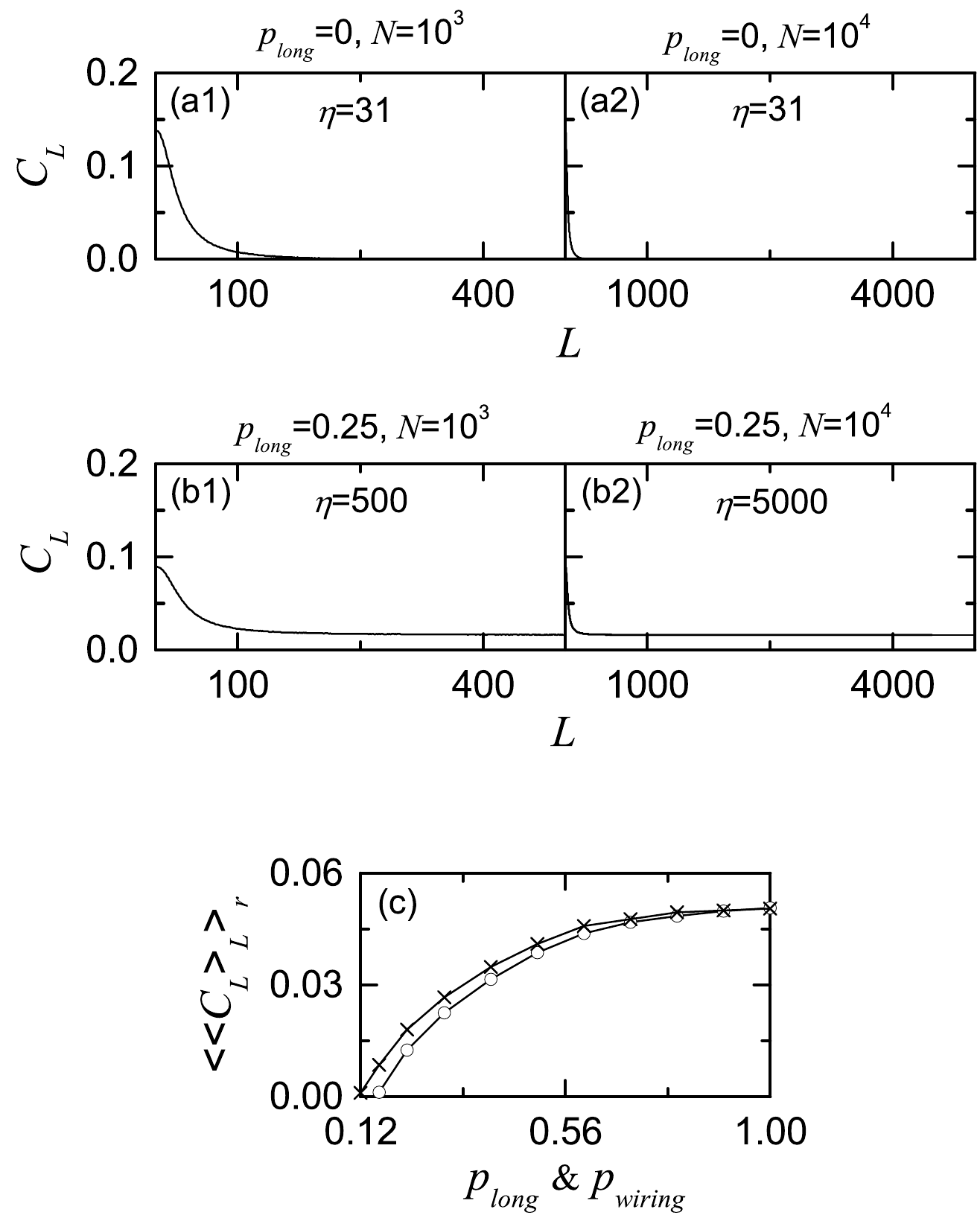}
\caption{Characterization of desynchronization-synchronization transition in terms of spatial correlations $C_L$. Unsynchronized state for $p_{long} = 0$: plots of the spatial correlation function $C_L$ between neuronal pairs versus the spatial distance $L$ for $N=$ (a1) $10^3$ and (a2) $10^4$. Synchronized state for $p_{long}$ = 0.25: plots of the spatial correlation function $C_L$ between neuronal pairs versus the spatial distance $L$ for $N=$ (b1) $10^3$ and (b2) $10^4$. (c) Plots of the average spatial correlation degree $\left< \left< C_L \right>_L \right>_r$ versus $p_{long}$ and $p_{wiring}$ for $N=10^3$ in the inhomogeneous (open circles) and the homogeneous
Watts-Strogatz (crosses) SWNs. The number of data used for the calculation of each temporal cross-correlation function $C_{i,j}(\tau)$ (the values at the zero-time lag ($\tau=0$) are used for calculation of $C_L$) is $2 \times 10^4$. $\left< \left< \cdots \right>_L \right>_r$ denotes double averaging over all lengths and 20 realizations.
}
\label{fig:Corr}
\end{figure}

In this subsection, we investigate the effect of network topology on emergence of sparsely synchronized rhythms by varying $p_{long}$. We first consider the population state for $p_{long} = 0$. As shown in Figs.~\ref{fig:PS}(a1) and \ref{fig:PS}(a2) for $N = 10^3$, the raster plot shows a zigzag pattern intermingled with inclined partial stripes of spikes with diverse inclinations and widths, and the IPSR kernel estimate $R(t)$ of Eq.~(\ref{eq:IPSR}) is composed of irregular parts with fluctuating amplitudes. For $p_{long} = 0$, the clustering coefficient $C$ in Fig.~\ref{fig:NT}(a1) is high, and hence partial stripes (indicating local clustering of spikes) seem to appear in the raster plot of spikes. As $N$ is increased to $10^4$, partial stripes become more inclined from the vertical [see Fig.~\ref{fig:PS}(b1)], and then spikes become more difficult to keep pace with each other. For this case, $R(t)$ shows noisy fluctuations with smaller amplitudes, as shown in Fig.~\ref{fig:PS}(b2). Consequently, the population state for $p_{long} = 0$ seems to be unsynchronized because $R(t)$ tends to be nearly stationary as $N$ increases to the infinity. As $p_{long}$ is increased from 0, LR connections begin to appear non-uniformly from LR interneurons. Due to the effect of these LR connections, the average path length $L_p$ becomes shorter, and the betweenness centralization $C_b$ becomes smaller (i.e., the load of communication traffic becomes less concentrated on LR interneurons due to decrease in the average betweenness centrality $\left< B \right>_{LR}$ of LR interneurons) [see Figs.~\ref{fig:NT}(b1), \ref{fig:BT}(e), and \ref{fig:BT}(f)]. Hence, with increasing $p_{long}$ efficiency of global communication between interneurons becomes better. Eventually, when passing a critical value $p_{long}^{(c)} \simeq 0.16$, synchronized population state emerges because of sufficient efficiency of information transfer between distant interneurons, which will be discussed in more detail in Figs.~\ref{fig:Order} and \ref{fig:Corr}. As an example of a synchronized state, we consider the case of $p_{long} = 0.25$. For $N = 10^3$, the degree of zigzagness for partial stripes in the raster plot is much reduced, and hence $R(t)$ shows a regular oscillation, as shown in Figs.~\ref{fig:PS}(c1)-\ref{fig:PS}(c2). Its amplitudes are much larger than those for $p_{long} = 0$, although there is a little variation in the amplitude. As $N$ is increased to $N = 10^4$, stripes become more vertically aligned, in contrast to the case of $N = 10^3$ [see Fig.~\ref{fig:PS}(d1)]. Hence, $R(t)$ shows more regular oscillations, and the amplitudes in each oscillating cycle are nearly the same, as shown in Fig.~\ref{fig:PS}(d2). Consequently, the population state for $p_{long} = 0.25$ seems to be synchronized because $R(t)$ tends to show regular oscillations as $N$ goes to the infinity. For this case, the population frequency $f_p$ of $R(t)$ is about 139 Hz [see Fig.~\ref{fig:PS}(e)]. For individual behaviors of spiking interneurons, histograms for interspike intervals (ISIs) and mean firing rates (MFRs) of the LR (denoted by circles) and the SR (represented by crosses) interneurons are also shown in Figs.~\ref{fig:PS}(f) and \ref{fig:PS}(g), respectively. Both ISI histograms of the LR and the SR interneurons have multiple peaks appearing at multiples of the global period $T_G (\simeq 7.2$ ms) of $R(t)$, because LR and SR interneurons exhibit intermittent spikings phase-locked to $R(t)$ at random multiples of the global period $T_G$ of $R(t)$.  Similar skipping phenomena of spikings (characterized with multi-peaked ISI histograms) have also been found in networks of coupled inhibitory neurons in the presence of noise where noise-induced hopping from one cluster to another one occurs \cite{GR}, in single noisy neuron models exhibiting stochastic resonance due to a weak periodic external force \cite{Longtin1,Longtin2}, and in inhibitory networks of coupled subthreshold neurons showing stochastic spiking coherence \cite{Kim1,Kim2}. Due to this type of stochastic spike skippings (i.e., random phase lockings), partial occupation occurs in the stripes of the raster plot of spikes (i.e., sparse stripes appear). Thus, the ensemble-averaged MFR $\left< f_i \right> (\simeq 31$ Hz) of individual LR and SR interneurons become much less than the population frequency $f_p (\simeq 139$ Hz), which results in the occurrence of sparse synchronization.

To determine the critical value for the desynchronization-synchronization transition, we employ a realistic thermodynamic order parameter $\cal{O}$ of Eq.~(\ref{eq:Order}), given by the mean square deviation of the IPSR kernel estimate $R(t)$. For a synchronized state, $R(t)$ exhibits an oscillation with a nonzero amplitude, while in the case of desynchronization $R(t)$ becomes nearly stationary, as shown in Fig.~\ref{fig:PS}.
Hence, the order parameter $\cal{O}$ approaches a nonzero (zero) limit value for the synchronized (unsynchronized) state in the thermodynamic limit of $N \rightarrow \infty$.
Figures \ref{fig:Order}(a) and \ref{fig:Order}(b) show plots of the order parameter $\cal{O}$ versus $p_{long}$ and $p_{wiring}$ for both cases of the inhomogeneous SWN and the Watts-Strogatz SWN, respectively. For $p_{long} < p_{long}^{(c)}$ ($p_{wiring} < p_{wiring}^{(c)}$), unsynchronized states exist because the values of $\cal{O}$ tends to zero as $N \rightarrow \infty$; $p_{long}^{(c)} \simeq 0.16$ and $p_{wiring}^{(c)} \simeq 0.12$.
As $p_{long}$ ($p_{wiring}$) passes the critical value $p_{long}^{(c)}$ ($p_{wiring}^{(c)}$), a transition to synchronization occurs because $\cal{O}$ becomes saturated to a nonzero limit value for $N \geq 3 \times 10^3$. As explained in the subsection \ref{subsec:TC}, the effect of LR connections for the Watts-Strogatz SWN is stronger than that for the inhomogeneous SWN, and hence the desynchronization-synchronization transition occurs at a smaller critical value in the case of the Watts-Strogatz SWN.

We further investigate the effect of LR connections on the above desynchronization-synchronization transition in terms of the spatial cross-correlation function $C_L$ of Eq.~(\ref{eq:SCC}) between neuronal pairs separated by a spatial distance $L$ $(=1, \cdots, N/2)$, given through average of all the temporal cross-correlations between the instantaneous individual spike rate (IISR) $r_i(t)$ of Eq.~(\ref{eq:IISR}) and $r_{i+L}(t)$ $(i=1,...,N)$ at the zero-time lag \cite{SWN-Kim}. As an example of desynchronization, we consider the case of $p_{long}=0$. Figure \ref{fig:Corr}(a1) shows the spatial cross-correlation function $C_L$ versus $L$ for $N=10^3$. For this case, $C_L$ exhibits a direct exponential decay to zero, because the data are well fitted with an exponential function with a characteristic correlation length $\eta$, $C_L = A~e^{-L/\eta}$ $(A=0.16$ and $\eta =31$). Then, one can think that the whole system is composed of independent partially synchronized blocks of size $\eta$. To examine emergence of population synchronization, we increase the number of interneurons as $N = 10^4$. In this case, $C_L$ in Fig.~\ref{fig:Corr}(a2) also shows an exponential decay to zero with the same exponential function. Hence, the correlation length $\eta$ remains unchanged, although the system size is increased 10 times. As a criterion for occurrence of population synchronization, a normalized correlation length $\tilde{\eta}$ ($= \eta / N$), representing the ratio of the correlation length $\eta$ to the system size $N$, was introduced \cite{SWN-Kim}. As $N$ is increased, $\tilde{\eta}$ tends to zero. Hence, the relative size of partially synchronized blocks (when compared to the whole system size) tends to zero. Consequently, no global population synchronization occurs for the case of $p_{long}=0$. Next, we consider a synchronized case for $p_{long}=0.25$. For this case, $C_L$ shows a decay to a nonzero limit $(\simeq 0.016)$, independently of $N$, as shown in Figs.~\ref{fig:Corr}(b1) and \ref{fig:Corr}(b2) for $N=10^3$ and $10^4$, respectively. Since $C_L$ is nonzero in the whole range of $L$, the correlation length $\eta$ becomes $N/2$, covering the whole system (note that the maximal distance between interneurons is $N/2$ because of the ring architecture on which interneurons exist), in contrast to the unsynchronized case of $p_{long}=0$. In this way, with increasing $N$ the correlation length $\eta$ increases as $N/2$, and the normalized correlation length $\tilde{\eta}$ has a nonzero limit value, 1/2, in the thermodynamic limit of $N \rightarrow \infty$. Consequently, the whole network is composed of just one synchronized block, and global population synchronization occurs for $p_{long}=0.25$ because $\eta$ covers the whole system thanks to a sufficient number of LR connections. The degree of population synchronization may be measured in terms of the average spatial correlation degree $\left< C_L \right>_L$ of $C_L$ over all lengths $L$. Figure \ref{fig:Corr}(c) shows plots of $\left< C_L \right>_L$ versus $p_{long}$ and $p_{wiring}$ in the inhomogeneous SWN (denoted by circles) and the Watts-Strogatz SWN (represented by crosses). As the parameters $p_{long}$ and $p_{wiring}$ increase, the fraction of LR connections $F_{LR}$ increases linearly [see Fig.~\ref{fig:NT}(d)]. Due to the effect of these LR connections, with increasing $p_{long}$ and $p_{wiring}$, $\left< C_L \right>_L$, representing the degree of population synchronization, increases. However, the increasing rate of $\left< C_L \right>_L$ becomes slower, because the effect of LR connections on population synchronization does not increase in proportion to the number of LR connections (i.e., with increasing $p_{long}$ and $p_{wiring}$ the effect of LR connections decreases). We also note that the Watts-Strogatz SWN has more LR connections than the inhomogeneous SWN, as shown in Fig.~\ref{fig:NT}(d). Hence, $\left< C_L \right>_L$  for the case of the Watts-Strogatz SWN is larger than that for the inhomogeneous SWN. However, with increasing the parameters $p_{long}$ and $p_{wiring}$ the effect of LR connections decreases, and hence the differences of $\left< C_L \right>_L$ for both cases become reduced.

\begin{figure}
\includegraphics[width=0.8\columnwidth]{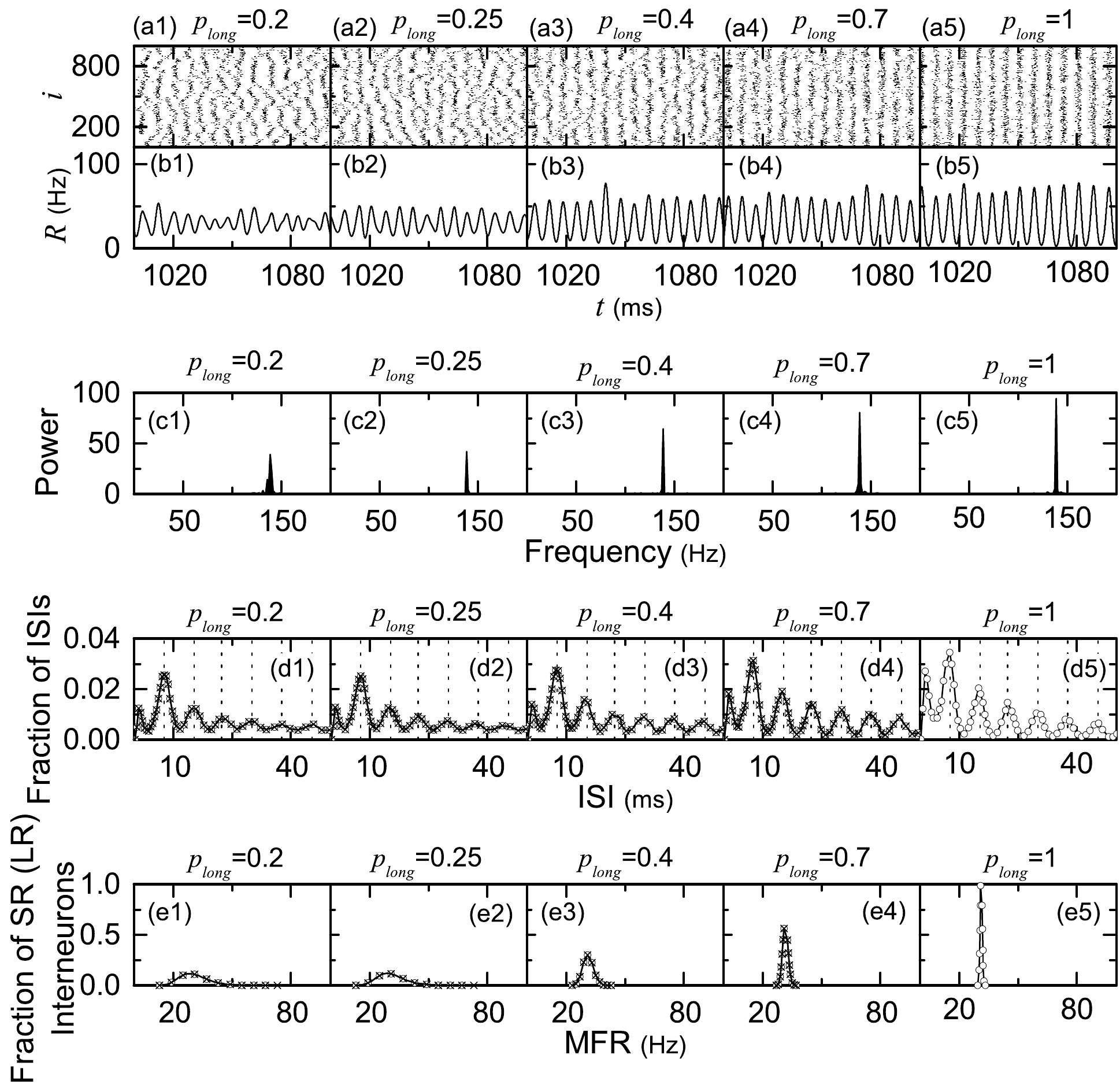}
\linespread{1.1}
\caption{Fast sparsely-synchronized rhythms for various values of $p_{long}$ in the inhomogeneous SWN with $N=10^3$. Raster plots of neural spikes and IPSR kernel estimates $R(t)$ for $p_{long}$= 0.2 [(a1) and (b1)], 0.25 [(a2) and (b2)], 0.4 [(a3) and (b3)], 0.7 [(a4) and (b4)], and 1.0 [(a5) and (b5)]. One-sided power spectrum of $\Delta R(t)$ $[=R(t) - \overline{R(t)}]$ (the overbar represents the time average) with mean-squared amplitude normalization for $p_{long}$= (c1) 0.2, (c2) 0.25, (c3) 0.4, (c4) 0.7, and (c5) 1.0. Each power spectrum is obtained from $2^{16} (=65536)$ data points. Interspike interval (ISI) histograms of the LR and the SR interneurons for $p_{long}$= (d1) 0.2, (d2) 0.25, (d3) 0.4, (d4) 0.7, and (d5) 1.0 (only LR interneurons exist). Open circles and crosses represent ISIs for the LR and the SR interneurons, respectively, and vertical dottd lines denote integer multiples of global period $T_G (\simeq 7.2$ ms) of $R(t)$. For each interneuron, $5 \times 10^4$ ISIs are obtained through 50 realizations. Hence, the ISI histograms are composed of $5 \times 10^7$ ISIs, and the bin size for the histograms is 0.5 ms. Histograms for the mean firing rates (MFRs) of individual SR and LR interneurons for $p_{long}=$ (e1) 0.2, (e2) 0.25, (e3) 0.4, (e4) 0.7, and (e5) 1.0 (only LR interneurons exist). Open circles and crosses represent MFRs for the LR and the SR interneurons, respectively. Averaging time for the MFR of each interneuron is $10^4$ ms in each realization, it is obtained via 50 realizations, and the bin size for the histogram is 0.5 Hz.
}
\label{fig:Sync}
\end{figure}

For various values of $p_{long} > p_{long}^{(c)}$, we study fast sparsely synchronized rhythms via comparison of their population behaviors with individual behaviors. As $p_{long}$ is increased, the zigzagness degree of partial stripes in the raster plots of spikes becomes reduced, as shown in Figs.~\ref{fig:Sync}(a1)-\ref{fig:Sync}(a5), and hence the pacing degrees between spikes in the stripes become increased. Consequently, with increasing $p_{long}$ the amplitudes of the IPSR kernel estimate $R(t)$ increase [see Figs.~\ref{fig:Sync}(b1)-\ref{fig:Sync}(b5)], which implies increase in the degree of population synchronization. Figures \ref{fig:Sync}(c1)-\ref{fig:Sync}(c5) show power spectra of $\Delta R(t)[=R(t) - \overline{R(t)}]$ (the overbar denotes the time average). For all values of $p_{long}$, single peaks appear at the population frequency $f_p$ $(\simeq 139$ Hz). As $p_{long}$ is increased, the heights of the peaks increase, while their widths decrease, thanks to the increase in the synchronization degree. As a result, with increasing $p_{long}$ the IPSRs $R(t)$ show more and more regular oscillations with larger amplitudes at the population frequency $f_p$ $(\simeq 139$ Hz), corresponding to the ultrafast rhythms. In contrast to regular population rhythms, individual LR and SR interneurons exhibit stochastic and sparse discharges as Geiger counters. Figures \ref{fig:Sync}(d1)-\ref{fig:Sync}(d5) and Figures \ref{fig:Sync}(e1)-\ref{fig:Sync}(e5) show histograms for ISIs and MFRs of the LR (denoted by circles) and the SR (represented by crosses) interneurons, respectively. Both the LR and the SR interneurons exhibit intermittent spikings phase-locked to the IPSR $R(t)$ at random multiples of the global period $T_G$ $(\simeq 7.2$ ms) of $R(t)$, and hence their ISI histograms have multiple peaks appearing at multiples of the global period $T_G$ of $R(t)$. Due to this kind of stochastic spike skipping (i.e., random phase lockings), sparse synchronization emerges (i.e., sparse stripes appear in the raster plot of spikes). Hence, the ensemble-averaged MFR $\left< f_i \right> (\simeq 31$ Hz) of LR and SR interneurons become much less than the population frequency $f_p (\simeq 139$ Hz), which implies that LR and SR interneurons make an average firing very sparsely about once during 4.5 global cycles. As $p_{long}$ is increased, ``valleys'' (corresponding to local minima) in the ISI histograms become lowered (i.e., they become deeper), and then multiple peaks have more clear shapes. (Conversely, with decreasing $p_{long}$ neighboring peaks overlap, and hence their shapes become less clear.) Accordingly, with increasing $p_{long}$ the widths of the MFR distributions become reduced, and the heights of the single peaks increase (i.e., they become sharper). In this way, as $p_{long}$ is increased stochastic spike skipping occurs in a more clear way.

\begin{figure}
\includegraphics[width=0.7\columnwidth]{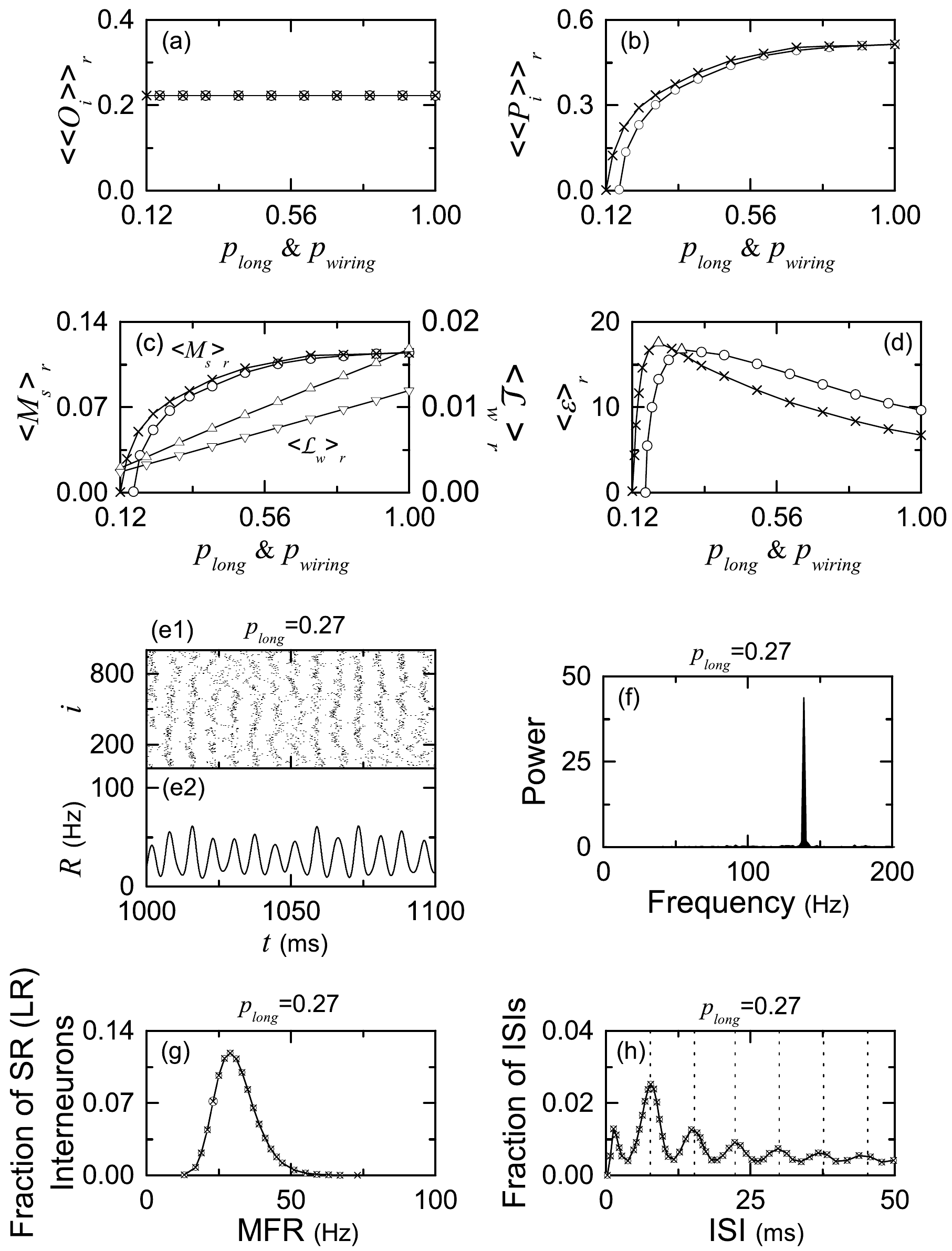}
\linespread{0.9}
\caption{Characterization of sparsely synchronized rhythms for $N=10^3$. Plots of (a) the average occupation degree $\left< \left< O_i \right> \right>_r$, (b) the average pacing degree $\left< \left< P_i \right> \right>_r$, (c) the statistical-mechanical spiking measure $\left< M_s \right>_r$ and the normalized wiring length $\left< {\cal{L}}_w \right>_r$, and (d) the dynamical efficiency $\left< \cal{E} \right>_r$ versus $p_{long}$ and $p_{wiring}$ in the inhomogeneous (circles and upper triangles) and the homogeneous Watts-Strogatz (crosses and lower triangles) SWNs. $\left< O_i \right>$, $\left< P_i \right>$, and $M_s$ are obtained by following the $3 \times 10^3$ stripes in the raster plot of spikes in each realization, and $\left< \cdots \right>_r$ denotes an average over 20 realizations. The optimal dynamical efficiencies are denoted by triangles. Optimal sparsely synchronized rhythm with the maximum dynamical efficiency $\cal{E}$ for $p_{long} = p_{long}^{(o)} (= 0.27)$: (e1) raster plot of neural spikes, (e2) IPSR kernel estimate $R(t)$, (f) one-sided power spectrum of $\Delta R(t) [= R(t) - \overline{R(t)}]$ (the overbar represents the time average) with mean-squared amplitude normalization, (g) histograms for the MFRs of the SR and the LR interneurons, and (h) histograms for the ISIs of the SR and the LR interneurons. The band width of the Gaussian kernel estimate for the IPSR $R(t)$ is 1 ms. The power spectrum is obtained from $2^{16} (=65536)$ data points. In (g) and (h), open circles and crosses represent those for the LR and SR interneurons, respectively. Averaging time for the MFR of each interneuron is $10^4$ ms in each realization, it is obtained via 50 realizations, and the bin size for the histogram is 0.5 Hz. For each interneuron, $5 \times 10^4$ ISIs are obtained through 50 realizations. Hence, the ISI histograms are composed of $5 \times 10^7$ ISIs, the bin size for the histograms is 0.5 ms, and vertical dotted lines denote integer multiples of global period $T_G (\simeq$ 7.2 ms) of $R(t)$.
}
\label{fig:Char}
\end{figure}

By varying $p_{long}$ in the whole range of fast sparsely synchronized rhythms, we also measure the degree of fast sparse synchronization in terms of a realistic statistical-mechanical spiking measure $M_s$, which was developed in our recent work \cite{RM}. As shown in Figs.~\ref{fig:Sync}(a1)-\ref{fig:Sync}(a5), population spike synchronization may be well visualized in a raster plot of spikes. For a synchronized case, the raster plot is composed of partially-occupied stripes (indicating sparse synchronization). To measure the degree of the sparse synchronization seen in the raster plot, a statistical-mechanical spiking measure $M_s$ of Eq.~(\ref{eq:SM}), based on the IPSR kernel estimate $R(t)$, was introduced by considering the occupation degrees of Eq.~(\ref{eq:OD})(representing the density of stripes) and the pacing degrees of Eq.~(\ref{eq:PD})(denoting the smearing of stripes) of the spikes in the stripes \cite{RM}: for more details, refer to the Appendix \ref{subsec:SMSM}. For each $p_{long}$, we follow $3 \times 10^3$ stripes and characterize sparse synchronization in terms of $\left< O_i \right>$ (average occupation degree), $\left< P_i \right>$ (average pacing degree), and the statistical-mechanical spiking measure $M_s$ for various values of $p_{long}$ in the sparsely synchronized region. The results (represented by circles) are shown in Figs.~\ref{fig:Char}(a)-\ref{fig:Char}(c), along with the results (denoted by crosses) in the Watts-Strogatz SWN. We note that the average occupation degree $\left< O_i \right>$ (denoting the average density of stripes in the raster plot) is nearly the same ($\left< O_i \right> \simeq 0.22$), independently of $p_{long}$. Hence, only a fraction (about 1/4.5) of the total interneurons fire in each stripe, which implies that individual interneurons fire about once during the 4.5 global cycles, agreeing well with the ensemble-averaged MFR $\left< f_i \right> (\simeq 31$ Hz). This partial occupation in the stripes results from stochastic spike skipping of individual interneurons which is seen well in the multi-peaked ISI histograms in Figs.~\ref{fig:Sync}(d1)-\ref{fig:Sync}(d5). Hence, the average occupation degree $\left< O_i \right>$ characterizes the sparseness degree of population synchronization well. On the other hand, as $p_{long}$ is increased, the average pacing degree $\left< P_i \right>$ (representing the average smearing of stripes in the raster plot) between spikes in the stripes increases due to appearance of LR connections affecting the efficiency of global communication between interneurons. However, with increasing $p_{long}$, the increasing rate for $\left< P_i \right>$ becomes slower because the effect of LR connections decreases, as in the case of the spatial correlation degree $\left< C_L \right>_L$ of Fig.~\ref{fig:Corr}(c). Figure \ref{fig:Char}(c) shows the statistical-mechanical spiking measure $M_s$ (taking into account both the occupation and the pacing degrees of spikes) versus $p_{long}$. Like the case of $\left< P_i \right>$, $M_s$ also makes an increase, because $\left< O_i \right>$ is nearly independent of $p_{long}$. $M_s$ is nearly equal to $\left< P_i \right>$ /4.5 due to sparse occupation [$\left< O_i \right> \simeq 0.22$]. The Watts-Strogatz SWN has more LR connections than the inhomogeneous SWN because LR connections for the Watts-Strogatz SWN appear via random uniform rewiring (which is made independently of the distance between interneurons), in contrast to the power-law connection probability (which decreases slowly with respect to the distance) for the inhomogeneous SWN. Thanks to the larger effect of these LR connections, the statistical-mechanical spiking measure $M_s$ of sparse synchronization in the Watts-Strogatz (denoted by crosses) is higher than that in the inhomogeneous SWN (represented by circles), as shown in Figs.~\ref{fig:Char}(a)-\ref{fig:Char}(c): the average pacing degree $\left< P_i \right>$ for the case of the Watts-Strogatz SWN is higher, although the occupation degree $\left< O_i \right>$ is the same (i.e., the ensemble-averaged MFRs are the same) for both cases. However, with increasing $p_{long}$ and $p_{wiring}$, the effect of LR connections decreases, and hence the difference in the synchronization degree for both cases becomes reduced. As the parameters $p_{long}$ and $p_{wiring}$ are increased from their critical values $p_{long}^{(c)} (\simeq 0.16)$ and $p_{wiring}^{(c)} (\simeq 0.12)$ in the inhomogeneous SWN and the Watts-Strogatz SWN, respectively, synchronization degree $M_s$ is increased because efficiency of global communication between interneurons becomes better. However, with increasing the parameters $p_{long}$ and $p_{wiring}$, the network axon wiring length becomes longer due to appearance of long-range short-cuts. Longer axonal connections are expensive because of material and energy costs. Hence, in view of dynamical efficiency we search for an optimal population rhythm emerging at a minimal wiring cost. An optimal fast sparsely synchronized rhythm may emerge via tradeoff between the synchronization degree and the wiring cost. The synchronization degree is given by the statistical-mechanical spiking measure $M_s$ shown in Fig.~\ref{fig:Char}(c), and the total wiring length in the whole network $L_w^{(total)}$ is given in Eq.~(\ref{eq:TWL}). We get a normalized wiring length ${\cal L}_w$ by dividing $L_w^{(total)}$ with $L_{total}^{(total,gl)}$ $[=\sum_{i=1}^{N} \sum_{j=1 (j \ne i)}^{N} l_w^{(ij)}]$ which is the total wiring length for the globally-coupled case:
\begin{equation}
{\cal L}_w= \frac{L_W^{(total)}}{L_W^{(total,gl)}}.
\end{equation}
A plot of ${\cal L}_w$ versus $p_{long}$ (denoted by upper triangles) is shown in Fig.~\ref{fig:Char}(c). It increases linearly with respect to $p_{long}$. Hence, with increasing $p_{long}$, the wiring cost becomes expensive. An optimal rhythm may emerge through tradeoff between the synchronization degree $M_s$ and the wiring cost ${\cal L}_w$. To this end, a dynamical efficiency $\cal{E}$ is given \cite{Buz2}:
\begin{equation}
 {\cal{E}}= \frac{\textrm{Synchronization\, Degree}~ (M_s)}{\textrm{Normalized\, Wiring\, Length}~ ({\cal L}_w)}.
\end{equation}
Figure \ref{fig:Char}(d) shows plot of $\cal{E}$ versus $p_{long}$. For $p_{long}=p^{(o)}_{long}$ $(\simeq 0.27)$, an optimal rhythm is found to emerge at a minimal wiring cost in an economic inhomogeneous SWN.
For the case of the Watts-Strogatz SWN, as the parameter $p_{wiring}$ is increased, the normalized wiring length ${\cal L}_w$ (denoted by lower triangles) increases more rapidly (i.e., the increasing rate is higher) than that in the inhomogeneous SWN, because the Watts-Strogatz SWN has more LR connections. Hence, an optimal fast sparsely synchronized rhythm for the Watts-Strogatz SWN appears at a smaller optimal value $p_{wiring}^{(o)} (\simeq 0.2)$. The dynamical efficiency $\cal{E}$ for $p_{wiring} > p_{wiring}^{(o)}$ decreases more rapidly than that for the inhomogeneous SWN because of more rapid increase in ${\cal L}_w$. An optimal fast sparsely synchronized rhythm for $p_{long}=0.27$ is shown in Fig.~\ref{fig:Char}(e1). Since the economic inhomogeneous SWN for the optimal case has a moderate clustering coefficient $C(p^{(o)}_{long})$ $(\simeq 0.32)$, the raster plot of spikes shows a zigzag pattern due to local clustering of spikes, and the IPSR kernel estimate $R(t)$ exhibits a regular ultrafast oscillation at a population frequency $f_p$ $(=139$ Hz) [see Figs.~\ref{fig:Char}(e2) and \ref{fig:Char}(f)]. In contrast to the population rhythm, individual LR and SR interneurons fire irregularly and sparsely with the ensemble-averaged MFR $\left< f_i \right> (\simeq 31$ Hz) [which is shown in Fig.~\ref{fig:Char}(g)] as Geiger counters, as shown well in the multi-peaked ISI histogram of Fig.~\ref{fig:Char}(h).

\section{Dynamical Responses to External Time-Periodic Stimuli}
\label{sec:DR}
In this section, we study the effect of the betweenness centralities $B$ of stimulated interneurons on the dynamical response to the external stimulus. Since $B$ characterizes the potentiality in controlling communication between other nodes in the rest of the network, effectiveness for transfer of stimulation effect in the whole network may vary depending on $B$. In the inhomogeneous SWN, LR interneurons have higher betweenness centralities than SR interneurons [see Fig.~\ref{fig:BT}(e)], in contrast to the homogeneous Watts-Strogatz SWN. Hence, it is expected that the degree of dynamical response may be larger for the case when the external stimuli are applied to LR interneurons with higher $B$.

We consider an optimal fast sparsely synchronized rhythm for $p_{long}^{(o)} =0.27$ in Fig.~\ref{fig:Char}. For this optimal case, we apply external time-periodic stimuli $S(t) [= A \sin(\omega_d t)]$ of Eq.~(\ref{eq:PD7}) to two sub-populations of LR and SR interneurons with different betweenness centralities, respectively and investigate their dynamical responses (suppression or enhancement) to $S(t)$. We also discuss the differences in their responses in relation to their betweenness centralities. Dynamical responses to such external periodic driving were studied in many works (e.g., in ensembles of bursting neurons with various types of coupling \cite{DR1,DR2,DR3}). Here, we set the driving angular frequency as $\omega_d (=2 \pi f_d) = 0.2$ rad/ms, corresponding to the ensemble-averaged MFR $\left< f_i \right> (\simeq 31$ Hz) of LR and SR interneurons [see Fig.~\ref{fig:Char}(g)]. Then, we investigate dynamical responses to $S(t)$ by varying the driving amplitude $A$.

\begin{figure}
\includegraphics[width=\columnwidth]{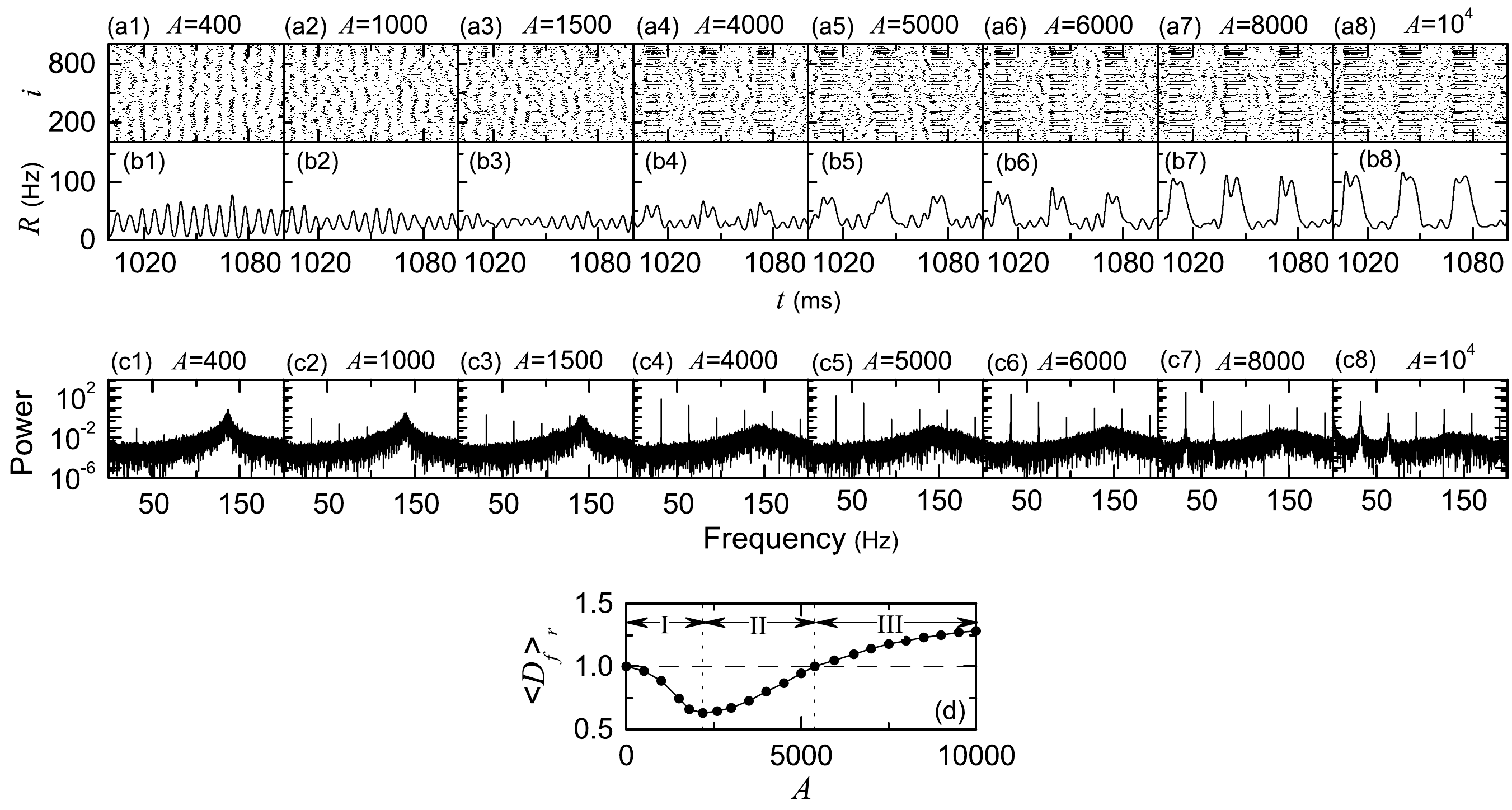}
\caption{Dynamical response for $p_{long}=0.27$ when an external time-periodic stimulus $S(t)$ is applied to 50 LR interneurons with higher betweenness centralities $B$ distributed near the average value $\left< B \right>_{LR} (=2711)$ of the LR interneurons. We vary the driving amplitude $A$ for a fixed driving angular frequency $\omega_d$ (=0.2 rad/ms). Raster plots of spikes and IPSR kernel estimates $R(t)$ for various values of $A$ are shown in (a1)-(a8) and (b1)-(b8), respectively. One-sided power spectra of $\Delta R(t) [= R(t) - \overline{R(t)}]$ (the overbar represents the time average) with mean-squared amplitude normalization are also given in (c1)-(c8). The band width of the Gaussian kernel estimate for each IPSR $R(t)$ is 1 ms, and each  power spectrum is obtained from $2^{16} (=65536)$ data points. (d) Plot of dynamical response factor $\left< D_f \right>_r$ (solid circles) versus $A$, where $I$, $II,$ and $III$ represent the 1st (synchronization suppression), 2nd (decrease in synchronization suppression), and 3rd (synchronization enhancement) stages, respectively. Here, $\left< \cdots \right>_r$ denotes an average over 20 realizations.
}
\label{fig:DRLR}
\end{figure}

We first apply $S(t)$ to $N_s (=50)$ selected LR interneurons which have their betweenness centralities $B$ in the range of $[2502,3043]$ near the ensemble-averaged betweenness centrality $\left< B \right>_{LR} (\simeq 2711)$ in the ensemble of LR interneurons. By varying $A$, we investigate dynamical responses to $S(t)$ for a fixed $\omega_d$ (=0.2). Here, we choose an appropriate number of stimulated interneurons as $N_s=50$: for small $N_s \ll 50$
the stimulation effect is very weak, and for $N_s \sim 50$ the stimulation effect becomes appreciable. Figures \ref{fig:DRLR}(a1)-\ref{fig:DRLR}(a8) show raster plots of spikes for various values of $A$. Their corresponding IPSR kernel estimates $R(t)$ are shown in Figs.~\ref{fig:DRLR}(b1)-\ref{fig:DRLR}(b8), and the power spectra of $\Delta R(t)$ $[= R(t) - \overline{R(t)}]$ (the overbar represents the time average) are also given in Figs.~\ref{fig:DRLR}(c1)-\ref{fig:DRLR}(c8). Then, the type and degree of dynamical response may be characterized in terms of a dynamical response factor $D_f$ \cite{DRF1,DRF2}:
\begin{equation}
D_f = \sqrt{\frac {Var(R_A)} {Var(R_0)}},
\label{eq:DRF}
\end{equation}
where $Var(R_A)$ and $Var(R_0)$ represent the variances of the IPSR kernel estimate $R(t)$ in the presence and absence of stimulus, respectively. If the dynamical response factor $D_f$ is less than 1, then synchronization suppression occurs; otherwise (i.e., $D_f >1$), synchronization enhancement takes place. Figure \ref{fig:DRLR}(d) shows a plot of $D_f$ versus $A$. Three stages are found to appear. Monotonic synchronization suppression (i.e., monotonic decrease in $D_f$ from 1), decrease in synchronization suppression (i.e., increase in $D_f$, but still $D_f<1$), and synchronization enhancement (i.e., increase in $D_f$ from 1) occur in the 1st (I) stage ($0< A < A^*_{1,LR}$), the 2nd (II) stage ($A^*_{1,LR} < A < A^*_{2,LR}$), and the 3rd (III) stage ($A>A^*_{2,LR}$), respectively; $A^*_{1,LR}\simeq 2186$ and $A^*_{2,LR} \simeq 5390$. Examples are given for various values of $A$; 1st stage ($A=400,$ 1000, and 1500), 2nd stage ($A=4000$ and 5000), and 3rd stage $(A=6000,$ 8000, and 10000).

In the 1st stage, sparse stripes (which exist originally for $A=0$) in the raster plot of spikes begin to break up due to the effect of the external AC stimulus $S(t)$. Since MFRs of stimulated interneurons increase, partially increased number of spikes appear at the sites of stimulated interneurons of the stripes, and then stripes begin to become partially horizontally scattered and merged to neighboring stripes. Many non-stimulated interneurons (i.e., major non-stimulated interneurons) which have synaptic connections with fast-firing stimulated interneurons fire slowly due to increased inhibition. On the other hand, a small number of non-stimulated interneurons (i.e. minor non-stimulated interneurons) which have no direct synaptic connections with stimulated interneurons receive synaptic inputs from major slowly-firing non-stimulated interneurons, and hence MFRs of minor non-stimulated interneurons become fast due to decreased inhibition. Consequently, the distribution of MFRs of non-stimulated interneurons becomes broadened (i.e., their dispersion increases). Due to the broadening of the MFR distribution, smearing and partial scattering occur at the sites of non-stimulated interneurons of the stripes in the raster plot of spikes. As $A$ is increased, the degree of this type of break-up of sparse stripes increases, as shown in Figs.~\ref{fig:DRLR}(a1)-\ref{fig:DRLR}(a3). In this 1st stage, with increasing $A$ the amplitudes of the corresponding IPSR kernel estimates $R(t)$ decrease because of break-up of sparse stripes [see Figs.~\ref{fig:DRLR}(b1)-\ref{fig:DRLR}(b3)]. Consequently, synchronization suppression occurs, which is well shown in the power spectra of Figs.~\ref{fig:DRLR}(c1)-\ref{fig:DRLR}(c3).  There appear
two kinds of peaks in the power spectra: one is associated with fast sparse synchronization and the other one is related to the external stimulus. Due to suppression of sparse synchronization, the height of the peak near 139 Hz (corresponding to fast sparse synchronization) begins to decrease, and its width tends to increase. Hereafter, this type of peak will be called the ``sparse-synchronization'' peak. Additional main peak and its harmonics, associated with the external driving of frequency $f_d (\simeq 31$ Hz), also begin to appear. These peaks will be referred to as ``stimulus'' peaks. With increasing $A$, the heights of these stimulus peaks increase due to the increased stimulus. In this way, the degree of sparse synchronization decreases due to a destructive effect of the external AC stimulus $S(t)$ (breaking up sparse stripes).

However, when passing a threshold $A^*_{1,LR}(\simeq 2186)$, a 2nd stage occurs where synchronization suppression stops and conversely synchronization degree begins to increase. For this case, the strength of the external AC stimulus becomes moderately increased and it begins to make a constructive effect. Stimulated interneurons begin to exhibit bursting-like firing behaviors in accordance to the external AC driving. As a result, partially scattered stripes of stimulated interneurons become more and more developed and ``bursting-like'' bands of spikes (without ``old'' sparse stripes) appear successively in accordance to the external AC stimulus frequency [e.g., see Figs.~\ref{fig:DRLR}(a4)-\ref{fig:DRLR}(a5)]. Hence, stimulated interneurons exhibit a kind of external phase lockings. With increasing $A$, density of spikes in the bursting-like bands increases. Within the bursting-like bands, spikes of non-stimulated interneurons are very sparse due to strong inhibition. When getting out of a bursting-like band, inhibition is much decreased, and hence a type of ``rebouncing effect'' occurs for the non-stimulated interneurons. Thus, a little clear stripes, composed of increased number of spikes of non-stimulated interneurons, appear between bursting-like bands: the degree of clearness of stripes tends to increase when approaching the neighboring bursting-like band. As a whole, under the moderate strength of external AC stimulus, both appearance of bursting-like bands of stimulated interneurons and occurrence of rebouncing effect outside the bands for non-stimulated interneurons lead to increase in synchronization degree (i.e., synchronization suppression stops and then synchronization degree begins to increase). The IPSR kernel estimates $R(t)$, reflecting the structure of the raster plots of spikes, are well shown in Figs.~\ref{fig:DRLR}(b4)-\ref{fig:DRLR}(b5). Large-amplitude oscillations appear in the regions of bursting-like bands: with increasing $A$, these amplitudes become larger. Due to the  rebouncing effect for the non-stimulated interneurons, small-amplitude oscillations also occur outside the regions of bursting-like bands. Mainly due to the bursting-like band effect, the synchronization degree is larger than that in the 1st stage. These behaviors are well shown in the power spectra of Figs.~\ref{fig:DRLR}(c4)-\ref{fig:DRLR}(c5). When compared with the 1st stage, stimulus peaks, associated with the external AC stimulus, become distinct. These clear stimulus peaks are related to formation of bursting-like bands (without ``original'' sparse stripes). Since no original sparse stripes exist in the bursting-like bands, the overall degree in sparse synchronization decreases, although sparse stripes appear outside the bursting-like bands via the rebouncing effect for the non-stimulated interneurons. Hence, the sparse-synchronization peak near 139 Hz becomes decreased and widened. Consequently, in the 2nd stage the external AC stimulus effect becomes dominant and leads to formation of bursting-like bands of stimulated interneurons via external phase lockings. Thanks to this constructive role of the external AC stimulus, synchronization suppression (occurring in the 1st stage) stops, and conversely degree in population synchronization begins to increase.

As $A$ is increased, the above tendency continues. Eventually, when passing another threshold $A^*_{2,LR} (\simeq 5390)$, the dynamical response factor $D_f$ becomes larger than unity, and hence a 3rd stage, where synchronization enhancement occurs, emerges. As shown in Figs.~\ref{fig:DRLR}(a6)-\ref{fig:DRLR}(a8), with increasing $A$, bursting-like bands become more distinct due to increased density of spikes. However, spikes of non-stimulated interneurons become much more sparse within the bursting-like bands because of increased strong inhibition. Hence, when getting out of a band, a decreased rebouncing effect occurs for the non-stimulated interneurons  and less clear (i.e., more smeared) stripes appear outside the bursting-like bands when compared with the 2nd stage. Since the effect of bursting-like bands is much more dominant, overall enhancement in population synchronization occurs in the 3rd stage. The IPSR kernel estimates $R(t),$ reflecting these structures in the raster plots of spikes, are shown well in Figs.~\ref{fig:DRLR}(b6)-\ref{fig:DRLR}(b8). As $A$ is increased, oscillations with larger amplitudes occur in the regions of  bursting-like bands. On the other hand, the amplitudes of oscillations outside the bands become smaller. Thanks to the dominant effect in the bursting-like bands, synchronization enhancement occurs. Due to the increased effect of bursting-like bands, stimulus peaks, associated with the external AC stimulus, become much more distinct. On the other hand, the sparse-synchronization peak near 139 Hz becomes more reduced, because no sparse stripes exist in the bursting-like bands and stripes outside the bands become smeared.

Based on the above results, it is found that the three stages appear through competition between the (original) fast sparse (mutual) synchronization and the external phase locking. Fast sparse synchronization breaks up gradually via occurrence of external phase lockings of stimulated interneurons. In the 1st stage, decrease in the sparse synchronization is dominant than the weak external phase lockings, and hence synchronization suppression occurs.
However, in the 2nd stage the external phase lockings begin to be dominant. Hence, the suppression of synchronization stops, and conversely synchronization degree begins to increase. Finally, in the 3rd stage the external phase lockings become much more dominant, and consequently synchronization enhancement appears.

\begin{figure}
\includegraphics[width=\columnwidth]{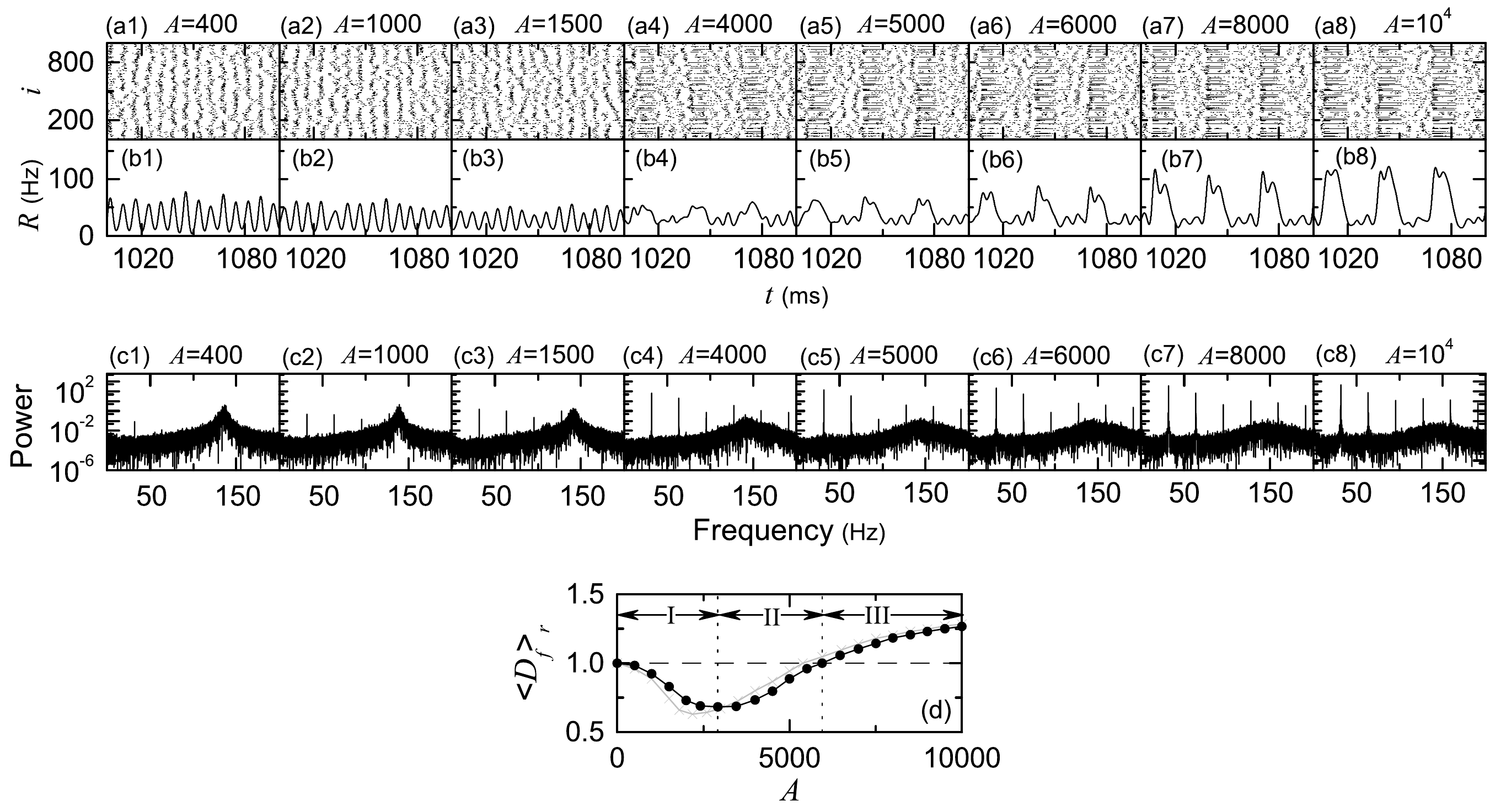}
\caption{Dynamical response for $p_{long}=0.27$ when an external time-periodic stimulus $S(t)$ is applied to 50 SR interneurons with lower betweenness centralities $B$ distributed near the average value $\left< B \right>_{SR} (=712)$ of the SR interneurons. We vary the driving amplitude $A$ for a fixed driving angular frequency $\omega_d$ (=0.2 rad/ms). Raster plots of spikes and IPSR kernel estimates $R(t)$ for various values of $A$ are shown in (a1)-(a8) and (b1)-(b8), respectively.  One-sided power spectra of $\Delta R(t) [ = R(t) - \overline{R(t)}]$ (the overbar represents the time average) with mean-squared amplitude normalization are also given in (c1)-(c8). The band width of the Gaussian kernel estimate for each IPSR $R(t)$ is 1 ms, and each  power spectrum is obtained from $2^{16} (=65536)$ data points. (d) Plot of dynamical response factor $D_f$ (solid circles) versus $A$, where $I$, $II,$ and $III$ represent the 1st (synchronization suppression), 2nd (decrease in synchronization suppression), and 3rd (synchronization enhancement) stages, respectively. For comparison, the dynamical response factor $D_f$ (gray line) for the case of stimulated LR interneurons in Fig.~\ref{fig:DRLR}(d) is also given. Here, $\left< \cdots \right>_r$ denotes an average over 20 realizations.
}
\label{fig:DRSR}
\end{figure}

Next, we apply the external AC stimulus $S(t)$ to $N_s (=50)$ selected SR interneurons which have their betweenness centralities $B$ in the range of $[697,733]$ near the ensemble-averaged betweenness centrality $\left< B \right>_{SR} (\simeq 712)$ in the ensemble of SR interneurons. We note that $\left< B \right>_{LR}$ is about 3.8 times as large as $\left< B \right>_{SR}$. Hence, transfer of stimulation effect from stimulated SR interneurons to the whole non-stimulated interneurons is expected to be less effective when compared with the case of stimulated LR interneurons. We investigate dynamical responses to $S(t)$  by varying $A$ for a fixed $\omega_d$ (=0.2). For various values of $A$, raster plots of spikes, IPSR kernel estimates $R(t)$, and the power spectra of $\Delta R(t)$ $[= R(t) - \overline{R(t)}]$ are shown in Figs.~\ref{fig:DRSR}(a1)-\ref{fig:DRSR}(a8),  Figs.~\ref{fig:DRSR}(b1)-\ref{fig:DRSR}(b8), and Figs.~\ref{fig:DRSR}(c1)-\ref{fig:DRSR}(c8), respectively. All of them are similar to those in Fig.~\ref{fig:DRLR} for the above ``LR-stimulated'' case.
Hence, the evolution of dynamical response to $S(t)$ follows the same three stages as in the LR-stimulated case. But, due to the difference in the effectiveness of transfer of stimulation effect, the degree of dynamical
response in each stage may be different, which can be quantitatively examined in terms of the dynamical response factor $D_f$. A plot of $D_f$ versus $A$ is shown in Fig.~\ref{fig:DRSR}(d). For comparison, $D_f$ for the LR-stimulated case (denoted by a gray line) is also given. Since the transfer of stimulation effect to the non-stimulated interneurons is less effective, change in the degree of dynamical response is less made in a slow way.
Hence, the threshold values, $A^*_{1,SR} (\simeq 2921)$ and $A^*_{2,SR} (\simeq 5942)$, for the 2nd and 3rd stages are larger than those for the LR-stimulated case (i.e., the 2nd and 3rd stages begin at larger threshold values due to the relatively slow evolution). Particularly, the degree of dynamical response (represented by $D_f$) is also reduced. In the 1st stage, the degree of synchronization suppression is less than that for the LR-stimulated case (i.e., less suppressed for the case of ``SR-stimulated'' case where the value of $D_f$ is larger than that for the LR-stimulated case). In the 2nd stage, the increasing rate for $D_f$ is larger for the LR-stimulated case, and when passing a threshold $A^*_{cr} (\simeq 3163),$ the gray line (representing $D_f$ for the LR-stimulated case) crosses the black line (denoting $D_f$ for the SR-stimulated case). Consequently, in the 2nd and 3rd stages for $A > A^*_{cr}$, $D_f$ for the SR-stimulated case becomes less than that for the LR-stimulated case (i.e., less enhanced for the SR-stimulated case). In this way, the degree of dynamical response (suppression or enhancement of population synchronization) is reduced because stimulated SR interneurons have lower betweenness centralities (leading to less effective transfer of of stimulation effect). For sufficiently large $A$, the population synchronization is governed mainly by the external phase lockings of stimulated interneurons (i.e., contributions of non-stimulated interneurons to population synchronization may be negligible), and hence the two black and gray curves of $D_f$ for both stimulated cases approach each other one (i.e., synchronization enhancements for both stimulated cases become very close for sufficiently large $A$).

\section{Summary}
\label{sec:SUM}
Instead of the homogeneous Watts-Strogatz SWN, we considered an inhomogeneous SWN composed of SR and LR interneurons by taking into consideration the inhomogeneous population of interneurons in real neural circuits. LR connections appear non-uniformly from the LR interneurons, in contrast to the case of the Watts-Strogatz SWN (where LR connections appear uniformly from all interneurons). For small $p_{long}$, LR interneurons have higher betweenness centralities (characterizing the potentiality in controlling communication between other interneurons) than SR interneurons, although they have the same average in- and out-degrees. Hence, the load of communication traffic is much concentrated on a few LR interneurons. However, with further increase in $p_{long}$, the number of LR interneurons increases, they share load of communication traffic, and hence the average betweenness centrality $\left< B \right>_{LR}$ of LR interneurons decreases [i.e., the (group) betweenness centralization $C_b$ also decreases]. As a result, the average path length $L_p$ becomes shorter, and the load of communication traffic is less concentrated on LR interneurons due to their decreased average betweenness centrality $\left< B \right>_{LR}$ (or equivalently due to decrease in the betweenness centralization $C_b$), which leads to better efficiency of global communication between interneurons.

We investigated the effect of the network topology (e.g., average path length and betweenness centrality) on emergence of sparsely synchronized rhythms with stochastic and intermittent neural discharges by varying $p_{long}$. As $p_{long}$ is increased, effective global communication between distant interneurons begins to become available due to appearance of LR connections from the LR interneurons. Eventually, when passing a small critical value $p_{long}^{(c)}$ $(\simeq 0.16)$, fast sparsely synchronized rhythms have been found to emerge because efficiency of global communication between interneurons becomes enough for occurrence of population synchronization thanks to shorter average path length $L_p$ and smaller average betweenness centrality $\left< B \right>_{LR}$ of LR interneurons. This transition to sparse synchronization has been well described via calculation of the realistic thermodynamic order parameter $\cal{O}$, based on the IPSR kernel estimate $R(t)$. For $p_{long} > p_{long}^{(c)}$, the IPSR kernel estimate $R(t)$ oscillates with the fast population frequency of 139 Hz, while individual SR and LR interneurons discharge spikes stochastically at low rates $(\sim 31$ Hz). Through calculation of the spatial correlation $C_L$, this type of global synchronization has been found to occur in the whole population because the spatial correlation length between neuronal pairs covers the whole system, thanks to sufficient number of LR connections. The degree of fast sparse synchronization has also been well measured in terms of the realistic statistical-mechanical spiking measure $M_s$. With increasing $p_{long}$ the degree of sparse synchronization increases, but its increasing rate becomes slower due to sufficient number of LR connections. By taking into consideration the axon wiring economy, an optimal sparsely synchronized rhythm has also been found to emerge at an optimal value $p_{long}^{(o)}$ $(\simeq 0.27)$ via trade-off between synchronization degree and wiring cost in an economic inhomogeneous SWN. These dynamical behaviors have also been compared with those in the Watts-Strogatz SWN, in connection with their network topologies. We note that the Watts-Strogatz SWN has more LR connections than the inhomogeneous SWN, because LR connections for the Watts-Strogatz SWN appear via random uniform rewiring (which is made independently of the distance between interneurons), in contrast to the power-law connection probability (decreasing slowly with respect to the distance) for the inhomogeneous SWN. Thanks to the larger effect of these LR connections, sparsely synchronized rhythms have been found to appear when passing a smaller critical value $p_{wiring}^{(c)} (\simeq 0.12)$, and the degree of sparse synchronization has also been found to become higher for the case of the Watts-Strogatz SWN. However, with increasing $p_{long}$ and $p_{wiring}$, the effect of LR connections decreases, and hence the difference in the synchronization degree for both cases becomes reduced.

Particularly, we note that the main difference between the inhomogeneous SWN and the homogeneous Watts-Strogatz SWN lies in the distributions of the betweenness centralities. Unlike the Watts-Strogatz SWN, dynamical responses to external AC stimulus vary depending on the type of stimulated interneurons (LR or SR interneurons) in the inhomogeneous SWN. For the optimal case of $p_{long}^{(o)} (\simeq 0.27)$, we considered two cases of external AC stimuli applied to sub-populations of LR and SR interneurons, respectively. Dynamical responses to these two cases of stimuli were investigated by varying the driving amplitude $A$ for a fixed driving angular frequency $\omega_d$ (=0.2), and they were discussed in relation to the betweenness centralities $B$ of stimulated interneurons. Three stages have thus been found to appear: monotonic synchronization suppression, decrease in synchronization suppression, and synchronization enhancement occur in the 1st, the 2nd, and the 3rd stages, respectively. The degree of dynamical response (such as synchronization suppression and enhancement) for the case of stimulated LR interneurons has been found to be larger (i.e., more suppressed or enhanced) than that for the case of stimulated SR interneurons, because stimulated LR interneurons with higher betweenness centralities make transfer of stimulation effect in the whole network more effectively than stimulated SR interneurons with lower betweenness centralities. In contrast, interneurons in the homogeneous Watts-Strogatz have nearly same betweenness centralities, and hence dynamical responses
to external stimuli have no particular dependence on randomly chosen interneurons. In this way, dynamical response to external stimuli are distinctly different for both cases of the inhomogeneous SWN and the Watts-Strogatz SWN, although the differences in their fast sparse mutual synchronization (in the absence of external stimuli) are not significantly large.

\begin{acknowledgments}
This research was supported by Basic Science Research Program through the National Research Foundation of Korea (NRF) funded by the Ministry of Education (Grant No. 2013057789).
\end{acknowledgments}

\appendix

\section{Methods for Characterization of Network Topology and Geometry}
\label{sec:NTG}
In this Appendix \ref{sec:NTG}, methods used for characterization of network topology and geometry are briefly explained.

\subsection{Clustering Coefficient}
\label{subsec:CC}
The clustering coefficient $C_i$ of a node $i$ in the network, representing the extent to which the neighborhood of the node $i$ is clustered, may be measured by the percentage of pairs of neighbors of the node $i$ that are also themselves neighbors \cite{CC}:
\begin{equation}
C_i = \frac{\rm the~number~of~triangles~in~the~network~with~the~node~{\it i}~as~one~vertex}{\rm the~number~of~all~possible~triangles~that~the~node~{\it i}~could~have~formed}.
\label{eq:CC}
\end{equation}
Then, the clustering coefficient $C$ in the whole network, denoting cliquishness of a typical neighborhood in the network, is given by the average of the clustering coefficients of all nodes \cite{SWN1}:
\begin{equation}
  C=\frac{1}{N} \sum_{i=1}^{N} C_i.
\end{equation}
We note that $C$ characterizes the local efficiency of information transfer: higher clustering effect occurs for large $C$.

\subsection{Average Path Length}
\label{subsec:AP}
The average path length $L_p$, representing typical separation between two nodes in the network, is given by the average number of connections between two nodes along the shortest path \cite{SWN1}:
\begin{equation}
L_p = \frac{1}{N(N-1)} \sum_{i=1}^{N} \sum_{j=1 (j\ne i)}^{N} l_p^{(ij)},
\label{eq:AP}
\end{equation}
where $l_p^{(ij)}$ is the shortest path length from the node $i$ to the node $j$ which may be easily obtained by employing the breadth-first searching algorithm
\cite{SP1,SP2}. We note that $L_p$ characterizes the global efficiency of information transfer between distant nodes: global efficiency is better for shorter $L_p$.

\subsection{Betweenness Centrality and Centralization}
\label{subsec:BC}
In the network science, centrality refers to indicators which identify the most important nodes within the network (i.e., the centrality indices are answers to the question ``which nodes are most
central?''). Historically first and conceptually simplest one is the degree centrality, which is defined by the number of edges of a node. This degree centrality represents the potentiality in communication activity.
Betweenness is also another centrality measure of a node within the network \cite{Bet1,Bet2}. Betweenness centrality of the node $i$ is given by the fraction of all the shortest paths between any two other nodes that pass
through the node $i$ \cite{Bet1,Bet2}:
\begin{equation}
B_i = \sum_{j=1 (j \ne i)}^{N} \sum_{k=1 (k \ne j \& k \ne i)}^{N} \frac{\sigma_{jk}(i)}{\sigma_{jk}},
\label{eq:BT}
\end{equation}
where $\sigma_{jk}(i)$ is the number of shortest paths from the node $j$ to the node $k$ passing through the node $i$, $\sigma_{jk}$ is the total number of shortest paths from the node $j$ to the node $k$, and
the number of shortest paths between two nodes may be easily obtained by using the breadth-first searching algorithm \cite{SP1,SP2}. This betweenness centrality $B_i$ characterizes the potentiality in controlling communication between other nodes in the rest of the network.
In the inhomogeneous SWN, LR interneurons are found to have higher betweenness centralities than SR interneurons. Hence, the load of communication traffic is concentrated on LR interneurons.
To examine how evenly the betweenness centrality is distributed among nodes (i.e., how evenly the load of communication traffic is distributed among nodes), we consider the group betweenness
centralization, denoting the degree to which the maximum betweenness centrality $B_{max}$ of the ``head'' LR interneuron exceeds the betweenness centralities of all the other interneurons.
The betweenness centralization $C_b$ is given by the sum of differences between the maximum betweenness centrality $B_{max}$ of the head LR interneuron and the betweenness centrality $B_i$ of other interneuron $i$
and normalized by dividing the sum of differences with its maximum possible value \cite{Bet1,Bet2}:
\begin{equation}
C_b = \frac{\sum_{i=1}^{N} (B_{\max}-B_i)}{\max \sum_{i=1}^{N} (B_{\max}-B_i)};
~~~ \max \sum_{i=1}^{N} (B_{\max}-B_i) = \frac{(N-1)(N^2 - 3N +2)}{2},
\label{eq:CB}
\end{equation}
where the maximum sum of differences in the denominator corresponds to that for the star network. Large $C_b$ implies that load of communication traffic is concentrated on the head LR interneuron,
and hence the head LR interneuron tends to become overloaded by the communication traffic passing through it. For this case, it becomes difficult to get efficient communication between
nodes due to destructive interference between so many signals passing through the head LR interneurons \cite{Bet3,SFN-Kim}. On the other hand, for small $C_b$ the load of traffic communication is less
concentrated on the head LR interneuron (i.e., it is more evenly distributed between nodes), and hence efficiency of communication between nodes becomes better.

\subsection{Wiring Length}
\label{subsec:WL}
As $p_{long}$ is increased, the network axon wiring length becomes longer due to increase in the LR connections. Longer axonal connections are expensive because of material and energy costs.
Hence, the (axon) wiring economy for the dynamical efficiency must be considered because wiring cost is an important constraint of the brain evolution \cite{Buz1,W_Review,Buz2,Sporns}.
Interneurons are equidistantly placed on a ring of radius $r$ $(=N / 2 \pi)$. Then, axonal wiring length $l_w^{(ij)}$ (i.e., geometrical wiring length of connection) between node $i$ and node $j$ is given by the arc length between two vertices $i$ and $j$ on the ring \cite{SWN-Kim,SFN-Kim}:
\begin{equation}
l_w^{(ij)} = \left\{ \begin{array}{l} |j-i| ~{\rm for}~ |j-i| \le \frac{N}{2} \\ N-|j-i| ~{\rm for}~ |j-i| > \frac{N}{2} \end{array}\right. .
\label{eq:WL}
\end{equation}
Then, the total wiring length in the whole network $L_w^{(total)}$ is given by:
\begin{equation}
L_w^{(total)}=\sum_{i=1}^{N} \sum_{j=1 (j \ne i)}^{N} a_{ij} l_w^{(ij)},
\label{eq:TWL}
\end{equation}
where $a_{ij}$ is the $ij$ element of the adjacency matrix $A$ of the network. The connection between vertices in the network is denoted by its $N \times N$ adjacency matrix $A(= \{ a_{ij} \})$
whose element values are 0 or 1. If $a_{ij} = 1$, then an edge from the vertex $i$ to the vertex $j$ exists; otherwise no such edges exists. This adjacency matrix $A$ corresponds to the
transpose of the connection weight matrix $W$ in Sec.~\ref{subsec:GE}.

We are also interested in the wiring lengths of outward connections from individual interneurons. The total wiring length $l_w^{(total)}(i)$ of the outward connections from the interneuron $i$ is given by:
\begin{equation}
l_w^{(total)}(i) = \sum_{j=1 (j \ne i)}^{N} a_{ij} l_w^{(ij)}.
\label{eq:TWLI}
\end{equation}
Then, the mean wiring length $\overline{l_w(i)}$ of the outward connections from the interneuron $i$ is:
\begin{equation}
\overline{l_w(i)}=\frac{l_w^{(total)}(i)}{N_c^{(i)}},
\label{eq:AWLI}
\end{equation}
where $N_c^{(i)}$ (=$\sum_{j=1 (j \ne i)}^{N} a_{ij}$) is the total number of outward connections from the interneuron $i$.
In the inhomogeneous SWN, the ensemble-averaged mean wiring length $\left< \overline{l_w(i)} \right>_{LR}$ of the outward connections in the ensemble of LR interneurons is longer than
$\left< \overline{l_w(i)} \right>_{SR}$ in the ensemble of SR interneurons, because LR connections appear non-uniformly from the LR interneurons.

\section{Methods for Characterization of Individual and Population Dynamics}
\label{sec:Dynamics}
In this Appendix \ref{sec:Dynamics}, methods used for characterization of individual and population dynamics are explained.

\subsection{Characterization of Individual Firing Behaviors}
\label{subsec:IB}
Firing behaviors of individual SR and LR interneurons are characterized in terms of the inter-spike interval (ISI) histogram and the mean firing rate (MFR) histogram \cite{Sparse1,Sparse2,Sparse3,Sparse4,Sparse5,Sparse6,SWN-Kim,SFN-Kim}. For each interneuron, $5 \times 10^4$ ISIs are collected through 50 realizations for the ISI histograms of the SR and the LR interneurons,
and the bin size for the histogram is 0.5 ms. The MFR for each SR or LR interneuron is calculated by following the membrane potential during the averaging time of $10^4$ ms after discarding the transient time of $10^3$ ms, and the bin size for the histogram is 0.5 Hz.

\subsection{Population Variables}
\label{subsec:PV}
In computational neuroscience, an ensemble-averaged global potential $V_G (t)$ in the whole population, containing $N$ FS Izhikevich interneurons,
\begin{equation}
V_G (t) = \frac {1} {N} \sum_{i=1}^{N} v_i(t)
\label{eq:GP}
\end{equation}
is often used for describing emergence of population neural synchronization in the whole population (e.g., sparse synchronization in a population of subthreshold neurons was described in terms of an ensemble-averaged global potential \cite{Kim1,Kim2}). However, to directly obtain $V_G(t)$ in real experiments is very difficult. To overcome this difficulty, instead of $V_G(t)$, we employ an experimentally-obtainable instantaneous population spike rates (IPSRs) which are often used as collective quantities showing population behaviors \cite{W_Review,Sparse1,Sparse2,Sparse3,Sparse4,Sparse5,Sparse6,RM,SWN-Kim}. The IPSR is obtained from the raster plot of neural spikes which is a collection of spike trains of individual interneurons. Such raster plots of spikes, where population spike synchronization may be well visualized, are fundamental data in experimental neuroscience. For the synchronous case, ``stripes" (composed of spikes and indicating population synchronization) are found to be formed in the raster plot. Hence, for a synchronous case, an oscillating IPSR appears, while for an unsynchronized case the IPSR is nearly stationary. To obtain a smooth IPSR, we employ the kernel density estimation (kernel smoother) \cite{Kernel}. Each spike in the raster plot is convoluted (or blurred) with a kernel function $K_h(t)$ to obtain a smooth estimate of IPSR $R(t)$:
\begin{equation}
R(t) = \frac{1}{N} \sum_{i=1}^{N} \sum_{s=1}^{n_i} K_h (t-t_{s}^{(i)}),
\label{eq:IPSR}
\end{equation}
where $t_{s}^{(i)}$ is the $s$th spiking time of the $i$th interneuron, $n_i$ is the total number of spikes for the $i$th interneuron, and we use a Gaussian kernel function of band width $h$:
\begin{equation}
K_h (t) = \frac{1}{\sqrt{2\pi}h} e^{-t^2 / 2h^2}, ~~~~ -\infty < t < \infty.
\label{eq:KF}
\end{equation}
Throughout the paper, the band width of the Gaussian kernel estimate is $h=1$ ms. Moreover, for the synchronous case, the population frequency $f_p$ of the regularly-oscillating IPSR kernel estimate $R(t)$ may be obtained from the one-sided power spectrum of $\Delta R(t)$ $[= R(t) - \overline{R(t)}]$ with the mean-squared amplitude normalization. The number of data for the power spectrum is $2^{16}(=65536)$, and the overline represents the time average.

\subsection{Thermodynamic Order Parameter}
\label{subsec:SWO}
As is well known, a conventional order parameter, based on the ensemble-averaged global potential $V_G(t)$, is often used for describing transition from synchronization to desynchronization in computational neuroscience \cite{Kim1,Kim2,Order1,Order2,Order3}. Recently, instead of the global potential, we used an experimentally-obtainable IPSR kernel estimate $R(t)$, and developed a realistic order parameter, which may be applicable in both the computational and the experimental neuroscience \cite{RM}. The mean square deviation of the IPSR kernel estimate $R(t)$,
\begin{equation}
{\cal{O}} \equiv \overline{(R(t) - \overline{R(t)})^2},
\label{eq:Order}
\end{equation}
plays the role of realistic order parameters ${\cal{O}}$ to determine a threshold for the synchronization-desynchronization transition, where the overbar represents the time average. Here, each order parameter is obtained through average over 20 realizations, and the averaging time for the calculation of the order parameter in each realization is $3 \times 10^4$ ms. Then, the order parameter ${\cal{O}}$, representing the time-averaged fluctuations of $R(t)$, approaches a nonzero (zero) limit value for the synchronized (unsynchronized) state in the thermodynamic limit of $N \rightarrow \infty$. This order parameter may be regarded as a thermodynamic measure because it concerns just the macroscopic IPSR kernel estimate $R(t)$ without any consideration between the macroscopic IPSR kernel estimate and microscopic individual spikes.

\subsection{Spatial Cross-Correlation Function}
\label{subsec:SCF}
To further understand the synchronization-desynchronization transition, we consider the ``microscopic'' dynamical cross-correlations between neuronal pairs \cite{SWN-Kim}. To get dynamical pair cross-correlations, each spike train of the $i$th neuron is convoluted with a Gaussian kernel function $K_h(t)$ of band width $h$ to get a smooth estimate of instantaneous individual spike rate (IISR) $r_i(t)$:
\begin{equation}
r_i(t) = \sum_{s=1}^{n_i} K_h (t-t_{s}^{(i)}),
\label{eq:IISR}
\end{equation}
where $t_{s}^{(i)}$ is the $s$th spiking time of the $i$th interneuron, $n_i$ is the total number of spikes for the $i$th interneuron, and $K_h(t)$ is given in Eq.~(\ref{eq:KF}). Then, the normalized temporal cross-correlation function $C_{i,j}(\tau)$ between the IISR kernel estimates $r_i(t)$ and $r_j(t)$ of the $(i,j)$ neuronal pair is given by:
\begin{equation}
C_{i,j}(\tau) = \frac{\overline{\Delta r_i(t+\tau) \Delta r_j(t)}}{\sqrt{\overline{\Delta r^2_i(t)}}\sqrt{\overline{\Delta r^2_j(t)}}},
\end{equation}
where $\Delta r_i(t) = r_i(t) - \overline{r_i(t)}$ and the overline denotes the time average.
Here, the number of data used for the calculation of each temporal cross-correlation function $C_{i,j}(\tau)$ is $2 \times 10^4$.
Then, the spatial cross-correlation function $C_L$ ($L=1,...,N/2)$ between neuronal pairs separated by a spatial distance $L$ is given through average of all the temporal cross-correlations between $r_i(t)$ and
$r_{i+L}(t)$ $(i=1,...,N)$ at the zero-time lag \cite{SWN-Kim}:
\begin{equation}
C_L = \frac{1}{N} \sum_{i=1}^{N} C_{i, i+L}(0) ~~~~ {\rm for~} L=1, \cdots, N/2.
\label{eq:SCC}
\end{equation}
Here, if $i+L > N$ in Eq.~(\ref{eq:SCC}), then $i+L-N$ is considered instead of $i+L$ because neurons lie on the ring. If the spatial cross-correlation function $C_L$ ($L=1,...,N/2)$ is nonzero in the whole range of $L$, then the spatial correlation length $\eta$ becomes $N/2$ (note that the maximal distance between interneurons is $N/2$ because of the ring architecture on which interneurons are placed) covering the whole network. For this case, global population synchronization emerges in the network; otherwise, desynchronization occurs.

\subsection{Satistical-Mechanical Spiking Measure}
\label{subsec:SMSM}
We measure the degree of sparse synchronization in terms of a realistic statistical-mechanical spiking measure, based on the IPSR kernel estimate $R(t)$ \cite{RM}. Spike synchronization may be well visualized in the raster plot of spikes. For a synchronized case, the raster plot is composed of partially-occupied stripes (indicating sparse synchronization), and the corresponding IPSR kernel estimate, $R(t)$, exhibits a regular oscillation. Each $i$th ($i=1,2,3,...$) global cycle of $R(t)$ begins from its left minimum, passes the central maximum, and ends at the right minimum [also, corresponding to the beginning point of the next $(i+1)$th global cycle]; the 1st global cycle of $R(t)$ appears after transient times of $10^3$ ms. Spikes which appear in the $i$th global cycle of $R(t)$ forms the $i$th stripe in the raster plot. To measure the degree of spike synchronization seen in the raster plot, a statistical-mechanical measure $M_s$, based on $R(t)$, was introduced by considering the occupation pattern and the pacing pattern of spikes in the stripes \cite{RM}. The spiking measures $M_i$ of the $i$th stripe [appearing in the $i$th global cycle of $R(t)$] is defined by the product of the occupation degree $O_i$ of spikes (representing the density of the $i$th stripe) and the  pacing degree $P_i$ of spikes (denoting the smearing of the $i$th stripe):
\begin{equation}
  M_i = O_i \cdot P_i.
\label{eq:SM}
\end{equation}
The occupation degrees $O_i$ in the $i$th stripe is given by the fractions of spiking interneurons in the $i$th stripe:
\begin{equation}
   O_i = \frac {N_{i}^{(s)}} {N},
\label{eq:OD}
\end{equation}
where $N_{i}^{(s)}$ is the number of spiking interneurons in the $i$th stripe.
For sparse synchronization with partially-occupied stripes, $O_i \ll 1$. The pacing degree $P_i$ of sparse spikes in the $i$th stripe can be determined in a statistical-mechanical way by taking into account their contributions to the macroscopic IPSR kernel estimate $R(t)$. An instantaneous global phase $\Phi(t)$ of $R(t)$ was introduced via linear interpolation in the two successive subregions forming global cycles \cite{RM}. The global phase $\Phi(t)$  between the left minimum (corresponding to the beginning point of the $i$th global cycle) and the central maximum is given by
\begin{equation}
\Phi(t) = 2\pi(i-3/2) + \pi \left(\frac{t-t_{i}^{(min)}}{t_{i}^{(max)}-t_{i}^{(min)}} \right) {\rm~~ for~} ~t_{i}^{(min)} \leq  t < t_{i}^{(max)},
\end{equation}
and $\Phi(t)$ between the central maximum and the right minimum [corresponding to the beginning point of the $(i+1)$th global cycle]
is given by
\begin{equation}
\Phi(t) = 2\pi(i-1) + \pi \left(\frac{t-t_{i}^{(max)}}{t_{i+1}^{(min)}-t_{i}^{(max)}} \right) {\rm~~ for~} ~t_{i}^{(max)} \leq  t < t_{i+1}^{(min)},
\end{equation}
where $t_{i}^{(min)}$ is the beginning time of the $i$th ($i=1, 2, 3, \cdots$) global cycle of $R(t)$ [i.e., the time at which the left minimum of $R(t)$ appears in the $i$th global cycle], and $t_{i}^{(max)}$ is the time at which the maximum of $R(t)$ appears in the $i$th global cycle. Then, the contributions of the $k$th microscopic spikes in the $i$th stripes occurring at the times $t_{k}^{(s)}$ to $R(t)$ is given by $\cos \Phi_k$, where $\Phi_k$  are the global phases at the $k$th spiking time [i.e., $\Phi_k \equiv \Phi(t_{k}^{(s)})$]. Microscopic spikes make the most constructive (in-phase) contributions to $R(t)$ when the corresponding global phases $\Phi_k$  is $2 \pi n$ ($n=0,1,2, \dots$), while they make the most destructive (anti-phase) contribution to $R(t)$ when $\Phi_k$ is $2 \pi (n-1/2)$. By averaging the contributions of all microscopic spikes in the $i$th stripes to $R(t)$, we obtain the pacing degrees $P_i$ of spikes in the $i$th stripe:
\begin{equation}
P_i = { \frac {1} {S_i}} \sum_{k=1}^{S_i} \cos \Phi_k
\label{eq:PD}
\end{equation}
where $S_i$ is the total number of microscopic spikes in the $i$th stripe. By averaging $M_i$ of Eq.~(\ref{eq:SM}) over a sufficiently large number $N_s$ of stripes, we obtain the statistical-mechanical spiking measure
$M_s$:
\begin{equation}
M_s =  {\frac {1} {N_s}} \sum_{i=1}^{N_s} M_i.
\label{eq:SM2}
\end{equation}
Here, we follow $3 \times 10^3$ global cycles in each realization, and obtain the average occupation degree, the average pacing degree, and the average statistical-mechanical spiking measure via average over 20 realizations.

\newpage
\begin{table}
\caption{Parameter values used in our computations; units of the capacitance, the potential, the current, the time, and the angular frequency are pF, mV, pA, ms, and rad/ms respectively.}
\label{tab:Parm}
\begin{ruledtabular}
\begin{tabular}{llllll}
(1) & \multicolumn{5}{l}{Single Izhikevich FS Interneurons \cite{Izhi3}} \\
& $C=20$ & $v_r=-55$ & $v_t=-40$ & $v_p=25$ & $v_b=-55$  \\
& $k=1$ & $a=0.2$ & $b=0.025$ & $c=-45$ & $d=0$  \\
\hline
(2) & \multicolumn{5}{l}{External Common Stimulus to Izhikevich FS Interneurons} \\
& $I_{DC} = 1500$ & $D=400$ \\
\hline
(3) & \multicolumn{5}{l}{Inhibitory GABAergic Synapse \cite{Sparse3}} \\
& $\tau_l=1$ & $\tau_r=0.5$ & $\tau_d=5$ & $V_{syn}=-80$ \\
\hline
(4) & \multicolumn{5}{l}{Synaptic Connections between Interneurons in Inhomogeneous and Homogeneous SWNs} \\
& $\sigma=20$ & $\kappa=100$ & \multicolumn{3}{l}{$\alpha=1$ (inhomogeneous SWN)} \\
& \multicolumn{5}{l}{$M_{syn}=50$ (homogeneous SWN)} \\
& \multicolumn{5}{l}{$J=1600$} \\
& \multicolumn{5}{l}{$p_{long}$ (inhomogeneous SWN) and $p_{wiring}$ (homogeneous SWN): Varying} \\
\hline
(5) & \multicolumn{5}{l}{External Time-Periodic Stimulus to LR and SR Interneurons} \\
& $\omega_d=0.2$ & $A:$ Varying
\end{tabular}
\end{ruledtabular}
\end{table}

\newpage

\newpage

\newpage

\newpage

\newpage

\newpage

\newpage

\newpage

\newpage

\newpage

\end{document}